\documentclass[a4paper,11pt]{article}
\pdfoutput=1
\usepackage{graphicx,rotating,hyperref,slashed,amsmath,xcolor,amssymb,amsfonts,colortbl,cite, subfigure,float}
\makeatletter 
\usepackage{amsmath}
\usepackage{graphics} 
\usepackage{graphicx} 
\graphicspath{{plots/}}
 
\hypersetup{colorlinks,bookmarksopen,bookmarksnumbered,
linkcolor=blus,pdfstartview=FitH,urlcolor=rossos,citecolor=verde}
\numberwithin{equation}{section}

\hypersetup{%
    ,urlcolor=rossos
    ,citecolor=rossos
    ,linkcolor=rossos
    }

\AtBeginDocument{
  \hypersetup{
    urlcolor=verdes,
    citecolor=verdes,
    linkcolor=verdes,
  }%
}

\allowdisplaybreaks

\def\lsim{\mathrel{\rlap{\lower3pt\hbox{\hskip0pt$\sim$}}
   \raise1pt\hbox{$<$}}}         
\def\gsim{\mathrel{\rlap{\lower4pt\hbox{\hskip1pt$\sim$}}
   \raise1pt\hbox{$>$}}}         

 \newcommand{\sfootnote}[1]{} 
\definecolor{bluc}{cmyk}{1,1,0,0.1}
\definecolor{rossoCP3}{cmyk}{0,.88,.77,.40}
\definecolor{rosso}{cmyk}{0,1,1,0.4}
\definecolor{rossos}{cmyk}{0,1,1,0.55}
\definecolor{rossoc}{cmyk}{0,1,1,0.2}
\definecolor{verdes}{cmyk}{0.92,0,0.59,0.4}

\hypersetup{colorlinks, bookmarksopen, bookmarksnumbered,
citecolor=verdes, linkcolor=bluc, pdfstartview=FitH, urlcolor=rossos}

\newcommand{\mio}[1]{}

\definecolor{Gray}{gray}{0.95}

\usepackage{multicol}
\usepackage{color}
\definecolor{rosso}{cmyk}{0,1,1,0.4}
\definecolor{rossos}{cmyk}{0,1,1,0.55}
\definecolor{rossoc}{cmyk}{0,1,1,0.2}
\definecolor{blu}{cmyk}{1,1,0,0.3}
\definecolor{blus}{cmyk}{1,1,0,0.6}
\definecolor{bluc}{cmyk}{1,1,0,0.1}
\definecolor{verde}{cmyk}{0.92,0,0.59,0.25}
\definecolor{verdec}{cmyk}{0.92,0,0.59,0.15}
\definecolor{verdes}{cmyk}{0.92,0,0.59,0.4}

\def\circa#1{\,\raise.3ex\hbox{$#1$\kern-.75em\lower1ex\hbox{$\sim$}}\,}

\newcommand{\beq}{\begin{equation}}
\newcommand{\eeq}{\end{equation}}

\newcommand{\bea}{\begin{eqnarray}}
\newcommand{\eea}{\end{eqnarray}}
\newcommand{\be}{\begin{equation}}
\newcommand{\ee}{\end{equation}}
\newfam\rsfsfam
\def\mathscr#1{{\fam\rsfsfam\relax#1}}

\def\circa#1{\,\raise.3ex\hbox{$#1$\kern-.75em\lower1ex\hbox{$\sim$}}\,}
\makeatletter

\def\hhref#1{\href{http://arxiv.org/abs/#1}{arXiv:#1}} 

\newcommand{\doi}[1]{\href{http://dx.doi.org/#1}{[doi]}}

\def\hhref#1{\href{http://arxiv.org/abs/#1}{arXiv:#1}} 
 
\def\art{\@ifnextchar[{\eart}{\oart}}
\def\eart[#1]#2#3#4#5#6{{\rm #2}, {\em #3 \bf #4} {\rm (#6) #5} ({\em #1})}

\def\article{\@ifnextchar[{\earticle}{\oarticle}}
\def\oarticle#1#2#3#4#5#6{{\rm #1}, {\em ``#6''}, {\rm #2 #3 (#5) #4}}
\def\earticle[#1]#2#3#4#5#6#7{{\rm #2}, {\em ``#7''}, {\rm #3 #4 (#6) #5}  [\hhref{#1}]}
\def\hepart[#1]#2{{\rm #2, \em#1}}
\def\heparticle[#1]#2#3{#2, {\em ``#3''} [\hhref{#1}]}

\newcounter{alphaequation}[equation]
\def\thealphaequation{\theequation\hbox to
0.6em{\hfil\alph{alphaequation}\hfil}}
\def\eqnsystem#1{
\def\@eqnnum{{\rm (\thealphaequation)}}
\def\@@eqncr{\let\@tempa\relax \ifcase\@eqcnt \def\@tempa{& & &} \or
  \def\@tempa{& &}\or \def\@tempa{&}\fi\@tempa
  \if@eqnsw\@eqnnum\refstepcounter{alphaequation}\fi
\global\@eqnswtrue\global\@eqcnt=0\cr}
\refstepcounter{equation} \let\@currentlabel\theequation \def\@tempb{#1}
\ifx\@tempb\empty\else\label{#1}\fi
\refstepcounter{alphaequation}
\let\@currentlabel\thealphaequation
\global\@eqnswtrue\global\@eqcnt=0 \tabskip\@centering\let\\=\@eqncr
$$\halign to \displaywidth\bgroup \@eqnsel\hskip\@centering
$\displaystyle\tabskip\z@{##}$&\global\@eqcnt\@ne
\hskip2\arraycolsep\hfil${##}$\hfil& \global\@eqcnt\tw@\hskip2\arraycolsep
$\displaystyle\tabskip\z@{##}$\hfil
\tabskip\@centering&\llap{##}\tabskip\z@\cr}
\def\endeqnsystem{\@@eqncr\egroup$$\global\@ignoretrue} \makeatother

\definecolor{fiorentina}{rgb}{.5,0,.5}

\def \bm#1{\mbox{\boldmath$#1$\unboldmath}}

\newcommand{\bn}{{\bf n}}

\newcommand{\fc}{\ensuremath{\mathcal{F}}}
\usepackage{cite}
\definecolor{rossoCP3}{cmyk}{0,.88,.77,.40}
\definecolor{rosso}{cmyk}{0,1,1,0.4}

\begin{document}

\vspace{1truecm}
 \begin{center}
\boldmath

{\textbf{\LARGE  
{ 
New test of modified gravity \\with gravitational wave experiments}}
}
\unboldmath
\end{center}
\unboldmath

\vspace{-0.2cm}

\begin{center}
\vspace{0.1truecm}

\renewcommand{\thefootnote}{\fnsymbol{footnote}}
\begin{center} 
{\fontsize{13}{30}
\selectfont  
N.~M.~Jim\'enez Cruz$^{a}$ \footnote{\texttt{nmjc1209.at.gmail.com}},
Flavio C.~S\'anchez$^{a,b}$\footnote{\texttt{flavio.sanchez@fisica.uaz.edu.mx}},
 Gianmassimo Tasinato$^{a, c}$ \footnote{\texttt{g.tasinato2208.at.gmail.com}}
} 
\end{center}


\begin{center}

\vskip 6pt
\textsl{$^a$ Physics Department, Swansea University, SA2 8PP, UK}
\\
\textsl{$^{b}$ 
Unidad Acad\'emica de F\'isica,
Universidad Aut\'onoma de Zacatecas,
98060, M\'exico
}
\\
\textsl{$^{c}$ Dipartimento di Fisica e Astronomia, Universit\`a di Bologna,\\
 INFN, Sezione di Bologna,  viale B. Pichat 6/2, 40127 Bologna,   Italy}
\vskip 4pt

\end{center}
\end{center}



\begin{abstract}
\noindent
We propose a new strategy to probe non-tensorial polarizations in the stochastic gravitational-wave (GW) background. Averaging over polarization angles, we find that three-point correlations of the GW signal vanish for tensor and vector modes, while scalar modes generically leave a nonzero imprint. This property makes the GW bispectrum a distinctive and robust diagnostic of scalar polarizations predicted in theories beyond General Relativity. We derive the corresponding response functions for ground-based interferometers, pulsar timing arrays, and astrometric observables, and we construct an optimal estimator together with simple Fisher forecasts for pulsar-timing sensitivity. As a proof of principle, we show that second-order GWs sourced by primordial magnetogenesis can be
characterized by large three-point
functions. Our results demonstrate that GW three-point correlations provide a novel observational window on physics beyond General Relativity.
\end{abstract}


\section{Introduction}

One of the most exciting opportunities offered by gravitational-wave (GW) observations is the ability to test General Relativity (GR) as a fundamental theory of gravity in regimes inaccessible to other experiments. See e.g. \cite{Maggiore:2007ulw,Maggiore:2018sht,Andersson:2019yve} for comprehensive 
textbooks on GW physics. Many alternatives to GR predict additional degrees of freedom, which may manifest as long-range interactions or as extra polarizations in GW signals \cite{Eardley:1973zuo,Eardley:1973br,Will:2018bme}. Such effects could be revealed either in signals from compact-binary mergers, or in the stochastic gravitational-wave background (SGWB, see e.g. \cite{Regimbau:2011rp,Romano:2016dpx,Caprini:2018mtu} for
reviews). The latter possibility is especially timely, given the recent evidence for a SGWB from pulsar timing arrays (PTAs) \cite{NANOGrav:2023gor,Reardon:2023gzh,Xu:2023wog,EPTA:2023fyk}.  

\smallskip

By combining data from multiple GW detectors, it is in principle possible to separate the  contributions of different polarization states to the SGWB. This idea has been extensively explored for ground-based interferometers, where signals can be linearly combined in the time domain to form so-called null streams, thereby isolating specific contributions from non-standard polarization modes:  see e.g. \cite{LIGOScientific:2018czr}
for recent  experimental bounds on extra
GW polarizations. In the context of pulsar timing arrays (PTAs) and astrometric experiments, the standard approach instead exploits the non-quadrupolar angular patterns that additional polarizations imprint on  PTA overlap reduction functions. By combining timing residuals from multiple pulsars and weighting them by appropriate powers of the noise covariance matrix, we can  extract possible contributions from extra polarizations (see, e.g., \cite{lee2008pulsar,Chamberlin:2011ev,yunes2013gravitational,gair2013testing,Gair:2015hra,Cornish:2017oic,arzoumanian2021nanograv,agazie2024nanograv,chen2024search,wu2022constraining,cornish2018constraining,gong2018polarizations,Mihaylov_2018,OBeirne:2018slh,Bernardo:2023zna,Liang:2023ary}). Intriguingly, recent analyses have even suggested tentative hints of possible modified-gravity effects \cite{Chen:2023uiz,NANOGrav:2023ygs,Agazie:2024qnx}. Nevertheless, isolating the signatures of different polarization states remains  challenging, due to both instrumental and astrophysical systematics. Imperfect knowledge of noise sources -- both intrinsic to the pulsar, or `local' uncertainties such as monopolar or dipolar contributions from clock or ephemeris errors  -- can mimic the effects of extra polarizations. Moreover, theoretical uncertainties complicate the problem: screening mechanisms may suppress the amplitude of additional modes relative to the standard tensorial (spin-2) ones \cite{Babichev:2013usa,Joyce:2014kja,Burrage:2017qrf}, while cosmological sources of SGWB may introduce nontrivial frequency dependencies, which further complicate attempts to disentangle among the distinct  polarization contributions, when combining
signals.

 \smallskip  

This motivates a natural question: \emph{is there a way to probe extra GW polarizations more directly,  without relying on combinations of signals?} In this work we propose a simple but powerful idea: the three-point correlation function of an isotropic and stationary SGWB provides a clean discriminator of scalar (spin-0) polarizations. We show that the response of GW detectors -- including ground-based interferometers, PTAs, and astrometric observatories -- to the GW \emph{three-point} function vanishes identically for tensor (spin-2) and vector (spin-1) modes, but is non-vanishing  for scalar modes. Thus, any detection of a non-zero  three-point function induced  
by an astrophysical or cosmological  source  of an isotropic SGWB would constitute an unambiguous and robust signature of scalar GW polarizations, uncontaminated by other degrees of freedom.  Moreover --  at least 
if noise source are in good approximations Gaussian -- this method can be less prone to certain  types of systematic errors (e.g. related with pulsar timing uncertainties), since noise contributions  do not directly contribute to the
three point function. 

\smallskip  

Our paper is organized as follows. In Section~\ref{sec_3ORF}, we prove that the isotropic three-point correlation function is sensitive only to GW scalar polarizations under well-defined assumptions, and we start discussing possible physical sources of three-point GW correlators. We then derive the three point response functions
of GW experiments to GW scalar polarizations.
 In Section~\ref{sec_opt}, we construct an optimal estimator to extract scalar contributions from GW data and present some preliminary forecasts for the sensitivity of PTA experiments. In Section~\ref{sec_cosex}, we illustrate our framework with an specific  early-universe scenario that can enhance the scalar GW three-point function to potentially observable levels. We summarize our findings in Section~\ref{sec_concl}. Throughout the paper we adopt natural units.

\section{Three-point functions and gravitational wave experiments:\\ a test for scalar polarizations}
\label{sec_3ORF}

\subsection{The idea}
\label{sec_idea}

We begin by providing an intuitive explanation of why measuring
  a non-vanishing three-point function of gravitational-wave (GW) signals would  necessarily indicate the presence of scalar GW polarizations
in a stochastic GW background (SGWB). In all our analysis, we  assume the SGWB background
to be stationary and isotropic, but it can be characterized by  non-Gaussian features. Moreover, additional
GW polarizations beside Einsteinian ones might be present in the data.

\smallskip

In the present section we focus on a noiseless GW  signal $s$, and defer a discussion
of the role of the noise to  Section~\ref{sec_opt}. The
 GW signal is schematically expressed as the contraction
\begin{equation}
s = D^{ij} h_{ij},
\end{equation}
where the tensor $D^{ij}$ encodes the geometry and response of the detector, while $h_{ij}$ denotes the GW field. Correlating such signals,
and averaging over all possible GW directions we can reveal the existence of an isotropic SGWB.
The GW field  can be Fourier decomposed into plane waves as
\begin{equation}
\label{eq:exphij}
h_{ij}(t,{\bf x}) = \sum_{\lambda=1}^6 \int_{-\infty}^{\infty} df \int d^2 \bn\, h_\lambda (f, \bn)\, {\bf e}^{(\lambda)}_{ij} (\bn) \, e^{i 2\pi f (t - \bn \cdot {\bf x})},
\end{equation}
with $h^*_\lambda (f, \bn) = h_\lambda (-f, \bn)$ ensuring that $h_{ij}$ is real. The index
$\lambda$ runs over the six possible polarizations in a general metric theory of gravity: $(+,\times)$ for tensor modes; $(v1,v2)$ for vector modes; and $(b,\ell)$ for the scalar breathing and longitudinal modes, respectively \cite{Eardley:1973zuo,Eardley:1973br,Will:2018bme}. In Eq.~\eqref{eq:exphij} we integrate
over all possible  directions of the unit vector $\bn$, as well as over frequencies which run on
the entire real line.

Introducing two orthonormal unit vectors $\bm{u}$ and $\bm{v}$ perpendicular to $\bn$, the six polarization tensors can be expressed as
\begin{subequations}\label{eq:PolTensors}
\begin{align}
\label{def_tpt}
{\bf e}^{(+)}_{ij} &= u_i u_j - v_i v_j, & \quad {\bf e}^{(\times)}_{ij} &= u_i v_j + v_i u_j, \\
\label{def_vpt}
{\bf e}^{(v1)}_{ij} &= n_i u_j + u_i n_j, & \quad {\bf e}^{(v2)}_{ij} &= n_i v_j + v_i n_j, \\
\label{def_spt}
{\bf e}^{(b)}_{ij} &= u_i u_j + v_i v_j, & \quad {\bf e}^{(\ell)}_{ij} &= n_i n_j.
\end{align}
\end{subequations}
The tensors in Eqs.~\eqref{def_tpt}, \eqref{def_vpt}, and \eqref{def_spt} correspond to spin-2, spin-1, and spin-0 polarizations, respectively, each with well-defined transformation properties under spatial rotations.  Our main interest is the scalar breathing mode $(b)$, and
we focus on it in this work. We do not consider   the longitudinal mode $(\ell)$ since in most Lorentz-covariant modifications of gravity such mode, if present, would correspond to a ghost degree of freedom, which would invalidate the setup and must
then be excluded.

An ambiguity arises from the freedom to choose the basis vectors $\bm{u}$ and $\bm{v}$. Any pair obtained by rotating through an angle $\psi$ in the plane orthogonal to $\bn$ defines an equally valid basis:
\begin{align}
\bm{u} &\to \bm{u}' = \cos\psi\, \bm{u} + \sin\psi\, \bm{v}, \nonumber \\
\bm{v} &\to \bm{v}' = -\sin\psi\, \bm{u} + \cos\psi\, \bm{v}, \label{rot_pola}
\end{align}
and the polarization tensors can be redefined accordingly. In fact, under this transformation, the spin-2 and spin-1 tensors rotate as
\begin{align}
\label{eq_trlp}
\begin{pmatrix}
\bf e^{(+)} \\
\bf e^{(\times)}
\end{pmatrix}
&\to
\begin{pmatrix}
\cos{2\psi} & -\sin{2\psi} \\
\sin{2\psi} & \cos{2\psi}
\end{pmatrix}
\begin{pmatrix}
\bf e^{(+)} \\
\bf e^{(\times)}
\end{pmatrix}, &
\begin{pmatrix}
{\bf e}^{(v1)} \\
{\bf e}^{(v2)}
\end{pmatrix}
&\to
\begin{pmatrix}
\cos{\psi} & -\sin{\psi} \\
\sin{\psi} & \cos{\psi}
\end{pmatrix}
\begin{pmatrix}
{\bf e}^{(v1)} \\
{\bf e}^{(v2)}
\end{pmatrix},
\end{align}
while the scalar polarization tensors ${\bf e}^{(b)}_{ij}$ and ${\bf e}^{(\ell)}_{ij}$ remain invariant.

Observable quantities derived from the GW signal $s$ should be independent of the arbitrary polarization angle $\psi$. As a concrete way to ensure this invariance, {\it we average over $\psi$ in the plane orthogonal to $\bn$} \cite{Belgacem:2024ohp} \footnote{Such choice of averaging has not been performed
in previous articles, as \cite{Powell:2019kid}.}.  This averaging has no effect on two-point functions, at least when focusing on a
isotropic stochastic gravitational wave background, as we
 do in this work. For instance, the combination
\begin{equation}
{\bf e}^{(+)}_{ij} {\bf e}^{(+)}_{ij} + {\bf e}^{(\times)}_{ij} {\bf e}^{(\times)}_{ij}
\end{equation}
is manifestly invariant under \eqref{eq_trlp} -- which leads to an overall factor $\cos^2{2 \psi}+\sin^2{2 \psi} =1$ -- so averaging over $\psi$ leaves it unchanged.  
In contrast, three-point functions involving only spin-2 and spin-1 polarizations contain terms linearly proportional to $\sin(n\psi)$ or $\cos(n\psi)$ (with $n$ equal to the spin), which vanish upon $\psi$-averaging. Only three-point functions involving spin-0 (scalar) polarizations survive this procedure.

\smallskip 
Therefore, within this approach, the detection of a non-zero three-point function in measurements of an isotropic SGWB would constitute a clear signature of scalar GW polarizations.  
Interestingly, this method is particularly direct, as it does not require constructing weighted  combinations of GW data to isolate specific polarizations in the SGWB.  
Moreover, being a null-type experiment,  it can reveal signatures of scalar polarizations even if their amplitude is suppressed with respect to Einsteinian tensor polarizations due to screening effects (see, e.g., \cite{Babichev:2013usa,Joyce:2014kja,Burrage:2017qrf} for reviews).  

\begin{figure}[t!]
    \centering
    \includegraphics[width=0.4\linewidth]{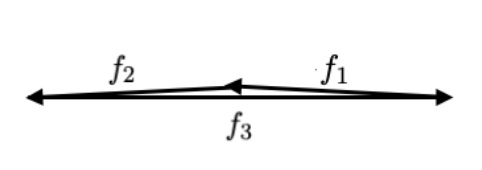}
        \includegraphics[width=0.4\linewidth]{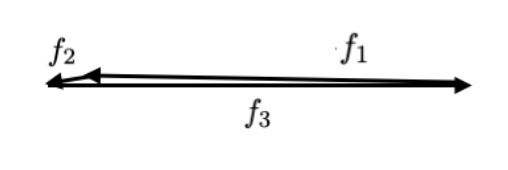}
    \caption{\small Representation of folded configurations for the 
    shape of the tensor bispectrum in momentum space. The three side
    of triangles lie on top of each other, and the length of the biggest triangle
    side, corresponding here to frequency $f_3$, is equal to the sum 
    of the other two sides, $f_1+f_2$.}
    \label{fig_fold}
\end{figure}

In  sections \ref{sec_warmup} and \ref{sec_pta}, we make these considerations more concrete by explicitly computing the overlap reduction functions for three-point correlations involving scalar modes. We initially assume the following structure for the two-point and three-point functions of the Fourier components of the GW modes

\begin{align}
\label{def_ps}
\langle h_{\lambda_1} (f_1, \bn_1)\, h_{\lambda_2} (f_2, \bn_2)\rangle 
&= \delta(f_1 + f_2)\, \frac{\delta^{(2)} (\bn_1 - \bn_2)}{4\pi}\, P_{\lambda_1 \lambda_2}(f_1), \\
\label{def_bis}
\langle h_{\lambda_1} (f_1, \bn_1)\, h_{\lambda_2} (f_2, \bn_2)\, h_{\lambda_3} (f_3, \bn_3) \rangle 
&= \delta(f_1 + f_2 + f_3)\, \frac{\delta^{(2)} (\bn_1 - \bn_3)}{4\pi}\, \frac{\delta^{(2)} (\bn_2 - \bn_3)}{4\pi}\, B_{\lambda_1 \lambda_2 \lambda_3}(f_1, f_2, f_3).
\end{align}
The function $P_{\lambda_1 \lambda_2}(f)$
is the power spectrum, while $B_{\lambda_1 \lambda_2 \lambda_3}$ is an instance of GW bispectrum.~\footnote{Other
possibilities can arise for the bispectrum, depending
on whether GW interactions involve derivatives -- more
on this in  the explicit setup of Section~\ref{sec_tpfgw}.}
The structure of Eq.~\eqref{def_ps} is consistent with statistical homogeneity and isotropy of the background. When writing a three-point function with the structure of \eqref{def_bis} --  besides imposing 
homogeneity and isotropy at the background level --  we further assume stationarity of the 
GW three-point function, following \cite{Powell:2019kid}. This assumption, though restrictive, is well motivated physically and greatly simplifies the analytic structure, making detector-response calculations tractable. The corresponding configuration in momentum space is that of a folded triangle, whose side length is given by the (absolute
value of) the frequencies $f_i$, and side directions are given by the vectors ${\bf n}_i$: in our case,
 two sides of the triangle are aligned, and lie on the third side. In fact, the three vectors ${\bf n}_1$, ${\bf n}_2$, ${\bf n}_3$ point in the same direction, thanks
to the structure of delta functions in Eq.~\eqref{def_bis}, and the sum of the length of two sides is equal to the length of the third one.  See Fig~\ref{fig_fold}. 

\medskip

Following Ref.~\cite{Powell:2019kid}, we now  show
that the Ansatz \eqref{def_bis} implies a stationary three-point function in real space. Recall that  we restrict  to the breathing scalar mode only.
 We define
\[
D(\bn) \equiv D^{ij} \, {\mathbf{e}}^{(b)}_{ij}(\bn)
\]
and, by Fourier transforming, we write the three-point function of the detector signal as  
\begin{align}
\langle s(t_1) s(t_2) s(t_3) \rangle
&= \int df_1 \, df_2 \, df_3 \int d^2 \bn_1 \, d^2 \bn_2 \, d^2 \bn_3 \;
\langle h_b(f_1, \bn_1) \, h_b(f_2, \bn_2) \, h_b(f_3, \bn_3) \rangle
\nonumber\\
&\quad \times
D(\bn_1) \, D(\bn_2) \, D(\bn_3) \;
e^{2\pi i \left(f_1 t_1 + f_2 t_2 + f_3 t_3\right)}
\, e^{-2\pi i \left(\bn_1 \cdot \mathbf{x}_1 + \bn_2 \cdot \mathbf{x}_2 + \bn_3 \cdot \mathbf{x}_3 \right)} .
\end{align}
Substituting Ansatz \eqref{def_bis} and performing the integrals over the delta functions, we find  
\begin{align}
\langle s(t_1) s(t_2) s(t_3) \rangle
&= \int df_1 \, df_2 \int d^2 \bn \;
{B}_{\bar h}(f_1, f_2,f_3) \, D^3(\hat n)
\, e^{2\pi i f_1 (t_1 - t_3) + 2\pi i f_2 (t_2 - t_3)}
\nonumber\\
&\quad \times
e^{-2\pi i \bn \cdot (\mathbf{x}_1 - \mathbf{x}_3) - 2\pi i \bn \cdot (\mathbf{x}_2 - \mathbf{x}_3)} ,
\label{eq_strpfun}
\end{align}
where the expression ${B}_{s}(f_1, f_2,f_3)$ denotes the bispectrum for the
scalar  breathing mode only~\footnote{In principle, mixed bispectra involving scalar, tensor, and vector modes can also be considered. However, any non-zero measurement of such quantities would already indicate the presence of scalar polarizations. We therefore focus on purely scalar bispectra.}, and recall that the frequencies in its argument are related
by the condition $f_1+f_2+f_3=0$.
It is now manifest that the three-point function is stationary, depending only on the time differences $t_1 - t_3$ and $t_2 - t_3$. As first discussed in \cite{Powell:2019kid}, stationary three-point functions can evade the arguments of \cite{Bartolo:2018rku}, which show that generic strain three-point functions, as in Eq.~\eqref{eq_strpfun}, are suppressed by decoherence effects between cosmological GW sources and detectors. In fact, we will return
on this point within an explicit example
that we develop in Section \ref{sec_cosex}, showing explicitly in a concrete setup
how dephasing works and how stationary
signals avoids it. Hence our hypothesis, besides simplifying  calculations, is
physically very relevant. Before continuing
to study the response functions
of GW experiments to GW 3-point functions,
we briefly discuss possible sources
for  GW non-Gaussianity.

\subsection{ Sources for the  GW three-point function}
\label{sec_revng}

Non-Gaussian features in the SGWB can arise from both astrophysical and cosmological sources, independently of whether modified gravity is considered. 


\smallskip

In the astrophysical case, the SGWB originates from the superposition of a large population of GW sources that are individually too weak to be resolved, but whose collective emission produces a measurable stochastic signal. If the sources are relatively rare and do not emit frequently, the signal exhibits characteristic ``popcorn'' features in the time domain, leading to non-Gaussian statistics \cite{Drasco:2002yd,Himemoto:2006hw,Seto:2009ju,Martellini:2014xia,Tsuneto:2018tif,Buscicchio:2022raf,Ballelli:2022bli}. Several methods have been proposed to capture such signatures, including extending likelihood analyses to incorporate non-Gaussian effects \cite{Drasco:2002yd,Martellini:2014xia} or searching for four-point correlators \cite{Seto:2009ju} (which, unlike the three-point correlators discussed here, can yield non-vanishing responses to spin-2 polarizations). See \cite{Romano:2016dpx} for a review. A related topic consists on analysing shot noise effects from discrete unresolved sources contributing to the astrophysical background. Such contributions are known to induce anisotropies (see e.g. \cite{Belgacem:2024ohp} for a recent account), and it would be interesting to explore their implications for non-Gaussianity, since the associated Poisson statistics naturally generates non-vanishing three-point cumulants.

\smallskip

In the cosmological case, any non-linearities in the generation of primordial GW can give rise to tensor non-Gaussianities. Numerous studies \cite{Adshead:2009bz,Dimastrogiovanni:2019bfl,Ricciardone:2017kre,Bartolo:2019oiq} have shown that cosmological mechanisms can produce a wide variety of non-Gaussian shapes, with amplitudes depending sensitively on the underlying scenario: see e.g. \cite{Bartolo:2018qqn} for a review. However, their detectability is severely limited: GW propagating over cosmological distances tend to lose correlations and thereby erase most non-Gaussian signatures \cite{Bartolo:2018rku}. Possible ways to circumvent this problem include: (i) searching for indirect effects of squeezed non-Gaussianities that modulate the GW power spectrum at small scales \cite{Dimastrogiovanni:2019bfl}, (ii) focusing on stationary non-Gaussian correlators \cite{Powell:2019kid} (this is the framework we adopt in this work; see Sec.~\ref{sec_idea}, as
well as Sec.~\ref{sec_cosex} for explicit examples), or (iii) studying non-Gaussianities in the anisotropies of the SGWB \cite{Bartolo:2019yeu,Bartolo:2019oiq}, which are not subject to such decorrelation effects.

\smallskip

We now turn to the detectability of stationary non-Gaussian features with GW experiments, and their potential as probes of modified gravity.

\subsection{A warm-up: the case
of coincident ground based detectors}
\label{sec_warmup}

Before turning to the study of overlap reduction functions for pulsar timing arrays and astrometry, we  consider here
a special case in the context  of ground-based detectors, which provides a simpler setting under certain assumptions. Our goal is to determine the response function of a triplet of coincident ground-based detectors to the GW three-point function
of Eq.~\eqref{def_bis}. To this end, we generalize the arguments of \cite{Allen:1997ad} (see also the textbook discussion in \cite{Maggiore:2007ulw}). A possible explicit realistic
example realising 
this configuration is the Einstein Telescope
detector \cite{Abac:2025saz}, if it will be built in a configuration at a 
single location.

We denote by $(1,2,3)$ the triplet of ground-based detectors. The signal measured say at detector $a$ is given by the contraction of the GW tensor with the detector tensor:
\begin{equation}
 s_1(t, {\bf x}) = h_{ij}(t, {\bf x})\, d_1^{ij}\,,
\end{equation}
where $d_1^{ij}$ encodes the detector characteristics (arm directions and lengths). For ground-based interferometers it can be expressed as
\begin{equation}
d_1^{ij} = X_1^i X_1^j - Y_1^i Y_1^j \,,
\end{equation}
with $X_1$ and $Y_1$ orthogonal unit vectors defining the arm directions.  Consequently, $d_1^{ii}=0$. Similar properties apply to detectors $2$ and $3$. 

As an illustrative example, let us consider the instrument response to spin-2 and spin-0 (breathing mode) polarizations. To start with, the two-point response functions are obtained from the signal correlations
\begin{eqnarray}
\langle s_1(t_1, {\bf x}_1)\, s_2(t_2, {\bf x}_2) \rangle &=&
\int d f \, e^{2 \pi f (t_1 - t_2)} \left[
\gamma^{\rm tens}_{12}({\bf x}_1,{\bf x}_2)\, I_T(f)
+ \gamma^{\rm sc}_{12}({\bf x}_1,{\bf x}_2)\, I_b(f)
\right] ,
\end{eqnarray}
where $I_T(f)$ and $I_b(f)$ are the intensities associated with tensor and breathing scalar polarizations, while
 the overlap response functions are given by 
\begin{eqnarray}
\gamma^{\rm tens}_{12}({\bf x}_1,{\bf x}_2) &=&
\sum_{\lambda=+,\times}\frac{d_1^{ij} d_2^{kl}}{2 \pi}
\int_0^{2 \pi} d \psi \int d^2 \hat n \,
e^{2 \pi i f\, {\bf n}\cdot({\bf x}_1-{\bf x}_2)}
\, {\bf e}_{ij}^\lambda({\bf n}) \, {\bf e}_{kl}^\lambda({\bf n}) \,, \\
\gamma^{\rm sc}_{12}({\bf x}_1,{\bf x}_2) &=&
\frac{d_1^{ij} d_2^{kl}}{2 \pi}
\int_0^{2 \pi} d \psi \int d^2 \hat n \,
e^{2 \pi i f\, {\bf n}\cdot({\bf x}_1-{\bf x}_2)}
\, {{\bf e}^{(b)}}_{ij}({\bf n}) \, {{\bf e}^{(b)}}_{kl}({\bf n}) \,.
\end{eqnarray}
As explained in Section~\ref{sec_idea}, we average over the polarization angle $\psi$ \cite{Belgacem:2024ohp}. For coincident detectors these expressions simplify considerably, assuming the arms are perpendicular, and the results can be written in closed form:
\begin{eqnarray}
    \gamma_{12}^{\rm tens} &=&\frac{8\pi}{5} \hskip1cm,\hskip1cm
    \gamma_{12}^{\rm sc} \,=\, \frac{4\pi}{15} \,.
    \label{res_2ptpt}
\end{eqnarray}

\medskip
In analogy, we  calculate the three-point response functions, which  are defined as
\begin{eqnarray}
\gamma^{\rm tens}_{123}({\bf x}_1,{\bf x}_2,{\bf x}_3) &=&
\sum_{\lambda=+,\times}\frac{d_1^{ij} d_2^{kl} d_3^{mn}}{2 \pi}
\int_0^{2 \pi} d \psi \int d^2 \hat n \,
e^{2 \pi i f\, {\bf n}\cdot({\bf x}_1-{\bf x}_2)}\,
e^{2 \pi i f\, {\bf n}\cdot({\bf x}_1-{\bf x}_3)} \,
{\bf e}_{ij}^\lambda({\bf n}) {\bf e}_{kl}^\lambda({\bf n}) {\bf e}_{mn}^\lambda({\bf n}) \,, 
\nonumber\\ \label{eq_3ten} \\[0.4em]
\gamma^{\rm sc}_{123}({\bf x}_1,{\bf x}_2,{\bf x}_3) &=&
\frac{d_1^{ij} d_2^{kl} d_3^{mn}}{2 \pi}
\int_0^{2 \pi} d \psi \int d^2 \hat n \,
e^{2 \pi i f\, {\bf n}\cdot({\bf x}_1-{\bf x}_2)}\,
e^{2 \pi i f\, {\bf n}\cdot({\bf x}_1-{\bf x}_3)} \,
{\bf e}^{(b)}_{ij}({\bf n}) {\bf e}^{(b)}_{kl}({\bf n}) {\bf e}^{(b)}_{mn}({\bf n}) \,. 
\nonumber\\
\label{eq_3sc}
\end{eqnarray}

To compute these expressions, we follow the method of \cite{Allen:1997ad}, focusing for simplicity on the idealized case of three coincident detectors (${\bf x}_1={\bf x}_2={\bf x}_3$), while allowing for different arm orientations. Equations~(\ref{eq_3ten}) and~(\ref{eq_3sc}) can then be written as
\begin{equation}
    \gamma^{ \sigma}_{123}({\bf x}_1,{\bf x}_2,{\bf x}_3) = d_1^{ij} d_2^{kl} d_3^{mn}\, \Gamma^{ \sigma}_{ijklmn} \,,
    \label{eq_3_gamma}
\end{equation}
where the superscript ``${ \sigma}$'' denotes either tensorial or scalar polarization.
The tensors $\Gamma^{\sigma}_{ijklmn}$ are symmetric under the exchanges $i \leftrightarrow j$, $k \leftrightarrow l$, $m \leftrightarrow n$, $ij \leftrightarrow kl$, $ij \leftrightarrow mn$, and $kl \leftrightarrow mn$, and are trace-free in each index pair. In the coincident-detector case, the six-index tensor can be constructed from Kronecker deltas: 
\begin{eqnarray}
 \Gamma_{ijklmn}^{ \sigma} &=&
A_1^{\sigma} \,\delta_{ij} \delta_{kl} \delta_{mn}
 \nonumber \\
&
+& A_2^{ \sigma} \left(\delta_{ik} \delta_{jl} \delta_{mn} + \delta_{jk} \delta_{il} \delta_{mn}
+ \delta_{im} \delta_{jn} \delta_{kl} + \delta_{in} \delta_{jm} \delta_{kl}
+ \delta_{km} \delta_{ln} \delta_{ij} + \delta_{kn} \delta_{ml} \delta_{ij}\right) \nonumber \\
&+& A_3^{ \sigma} \left(\delta_{ik} \delta_{jm} \delta_{ln} + \delta_{im} \delta_{ln} \delta_{jk}
+ \delta_{il} \delta_{jm} \delta_{kn} + \delta_{ik} \delta_{jn} \delta_{lm}\right) \, ,
    \label{eq_Gamm_gen}
\end{eqnarray}
where $A_i^{\sigma}$ $(i=1,2,3)$ are constants to be determined. Contracting Eq.~(\ref{eq_3_gamma}) with Eq.~(\ref{eq_Gamm_gen}), and exploiting
the tracelessness of $d_{ij}$,
we obtain
\begin{equation}
    \gamma_{123}^{\sigma} = 4 A_3^{\sigma} \, d_1^{ij} d_{2i}^{\ \ k} d_{3jk}\,.
\end{equation}


\noindent
Following then  step-by-step the approach of \cite{Allen:1997ad}, contracting Eq.~\eqref{eq_Gamm_gen} 
with the coefficients of the $A_i^{\sigma}$ and combining the results,  we 
arrive at,
\begin{eqnarray}
\gamma^{\rm tens}_{123} &=& 0 \,, \label{thrpt_tens}\\
\gamma^{\rm sc}_{123} &=& \frac{159 \pi ^2}{284}\, d_1^{ij} d_{2i}^{\ \ k} d_{3jk} \,.
\end{eqnarray}
Hence, as anticipated in Section~\ref{sec_idea}, while the three-point response function vanishes for tensor polarizations, it is nonzero for scalar polarizations and can therefore lead to  a distinctive observational handle on such signals. It is straightforward to show that, for the same reasons, three-point correlations to vector
polarizations vanish. 

\subsection{Overlap reduction functions for pulsar timing arrays and astrometry}
\label{sec_pta}

We now compute the three-point correlators relevant for pulsar timing array (PTA) and astrometry measurements, including the possibility of cross-correlating PTA and astrometry signals. Interestingly, the resulting overlap reduction functions exhibit a remarkably simple structure, enabling novel and direct tests of scalar polarizations with these experiments. In what follows we focus on tensor and breathing scalar polarizations, since the case of vector modes is straightforward.  

A large body of work has already investigated the response of PTA and astrometry to gravitational waves (GW), see e.g. \cite{Romano:2016dpx} for a review. 
 Our results extend these efforts by introducing three-point correlation functions into PTA and astrometry analyses, with the aim to extract
effects of GW scalar polarization.  

\smallskip
\noindent
{\bf Pulsar Timing Arrays.}  
For PTAs, we compute correlators of the GW-induced redshift in the pulse arrival rate, given by (see e.g.~\cite{Maggiore:2018sht} for a textbook discussion)
\begin{equation}
    z(t)\,=\,\frac{n^i n^j}{2\,(1+\mathbf{n}_a \cdot \mathbf{n})}
    \left[h_{ij}(t,\mathbf{x}=0)-h_{ij}(t-\tau_a,\mathbf{x}_a)\right],
\end{equation}
where $\tau_a$ is the light travel time between the Earth and pulsar $a$,  $\mathbf{n}_a$ is the unit vector pointing towards pulsar $a$, and $\mathbf{n}$ the unit vector indicating the GW propagation direction. 

We begin by recalling (without derivation) the known results for two-point correlators, following the approach of~\cite{Maggiore:2018sht}. Denoting again by $I_T(f)$ and $I_b(f)$ the intensities associated with tensor and breathing scalar polarizations, the ensemble average of two PTA measurements is
\begin{eqnarray}
    \langle z_a(t)\,z_b(t)\rangle 
    &=& \frac{1}{2}\int_{-\infty}^{\infty} df 
    \left[I_T(f)\,\kappa^{\text{tens}}_{ab}(\theta_{ab})
    +I_b(f)\,\kappa^{\text{scal}}_{ab}(\theta_{ab})\right],
\end{eqnarray}
where the overlap reduction functions $\kappa^{\text{tens}}$ and $\kappa^{\text{scal}}$ are obtained by integrating over all GW directions. Neglecting pulsar terms, the tensor overlap function reduces to the Hellings--Downs curve \cite{Hellings:1983fr},
\begin{eqnarray}
    \kappa^{\text{tens}}_{ab}(\theta_{ab}) = 
    x_{ab}\log x_{ab}-\tfrac{1}{6}x_{ab}+\tfrac{1}{3},
\end{eqnarray}
with 
\[
x_{ab}=\tfrac{1}{2}\big(1-\cos\theta_{ab}\big),
\]
where $\theta_{ab} = \text{arccos}{({\hat n}_a \cdot {\hat n}_b)}$ is the angular separation between pulsars $a$ and $b$.  
The scalar overlap function instead reads (see e.g.~\cite{Romano:2016dpx})
\begin{eqnarray}
    \kappa^{\text{scal}}_{ab}(\theta_{ab})=\tfrac{1}{6}(2-x_{ab}).
\end{eqnarray}

\medskip
We now extend this construction to higher-order correlators. The three-point correlator of PTA measurements is
\begin{eqnarray}
\langle z_a(t)\,z_b(t)\,z_c(t)\rangle
= \frac{1}{2}\int_{-\infty}^{\infty} df 
\Big[B_T(f)\,\kappa^{\text{tens}}_{abc}(\mathbf{n}_{a},\mathbf{n}_{b},\mathbf{n}_{c})
+ B_{b}(f)\,\kappa^{\text{scal}}_{abc}(\mathbf{n}_{a},\mathbf{n}_{b},\mathbf{n}_{c})\Big],
\end{eqnarray}
where $B_T$ and $B_{b}$ are the tensor and scalar bispectra (see Eq.~\eqref{def_bis}), and $\mathbf{n}_{a,b,c}$ denote the pulsar directions. The overlap functions are
\begin{eqnarray}
    \kappa^{\text{tens}}_{abc}(\mathbf{n}_{a},\mathbf{n}_{b},\mathbf{n}_{c}) 
    &=& \frac{1}{2\pi}\int_{0}^{2\pi} d\psi \int d^2\hat{n}\,
    \sum_{\lambda=+,\times} F_{a}^{\lambda}(\mathbf{n}_a)F_{b}^{\lambda}(\mathbf{n}_b)F_{c}^{\lambda}(\mathbf{n}_c), \\
    \kappa^{\text{scal}}_{abc}(\mathbf{n}_{a},\mathbf{n}_{b},\mathbf{n}_{c}) 
    &=& \frac{1}{2\pi}\int_{0}^{2\pi} d\psi \int d^2\hat{n}\,
    F^{\text{b}}(\mathbf{n}_a)F^{\text{b}}(\mathbf{n}_b)F^{\text{b}}(\mathbf{n}_c),
\end{eqnarray}
with antenna response functions
\begin{equation} \label{eq: PTA Corr Factor}
    F^{\sigma}(\mathbf{n}_k)=
    \frac{n^{i}_{k}n^{j}_{k}}{2\,(1+\mathbf{n}\cdot\mathbf{n}_k)}\,e_{ij}^{\sigma},
\end{equation}
for $\sigma=\{+,\times, \text{b}\}$, where $k$ labels the pulsar.  

\begin{figure}[t!]
    \centering
    \includegraphics[width=0.8\linewidth]{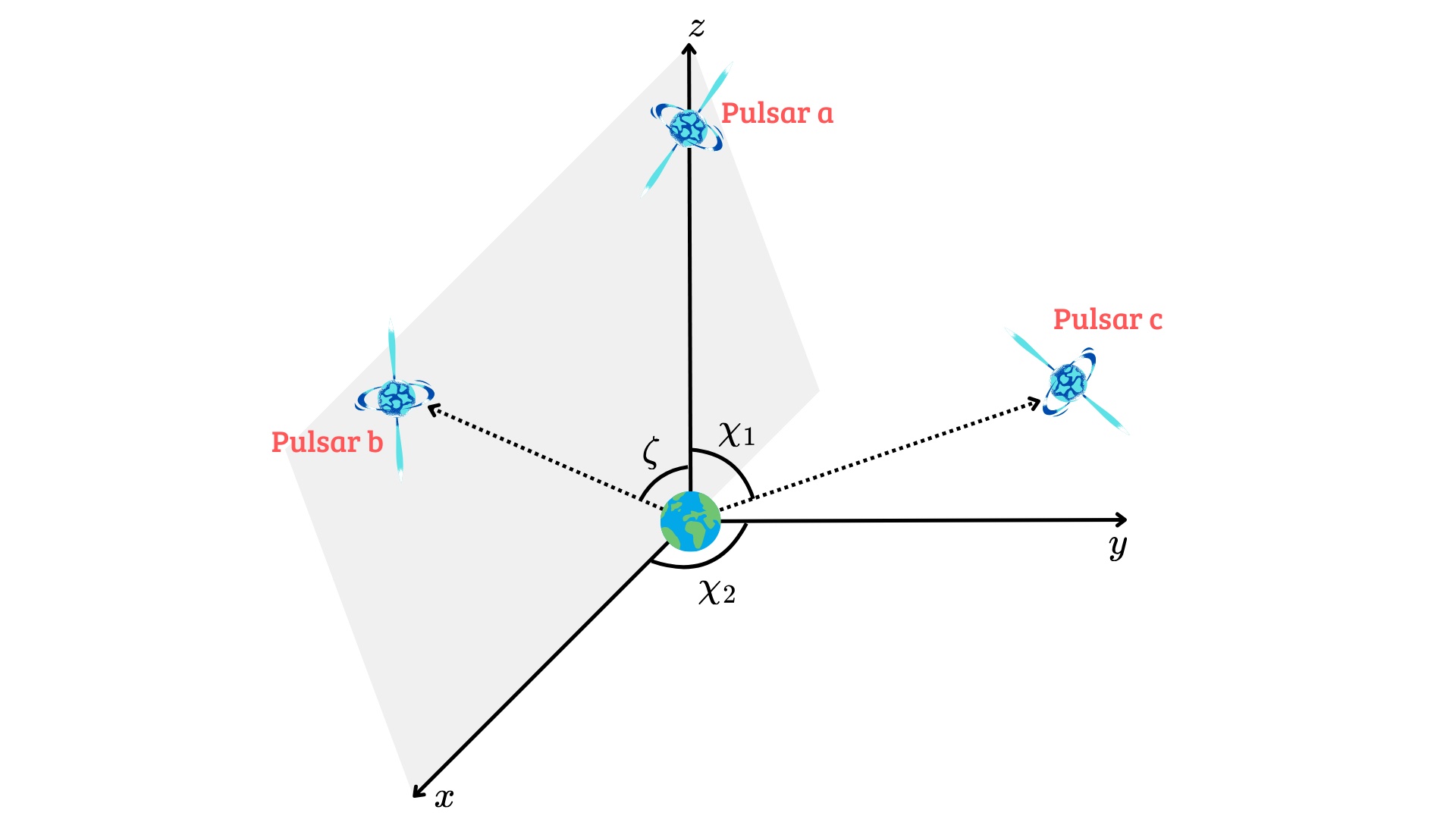}
    \caption{\small Geometry used to compute three-point overlap functions from pulsars $a$, $b$, and $c$. The Earth is at the origin, pulsar $a$ is placed on the $z$-axis, pulsar $b$ lies in the $xz$-plane, and pulsar $c$ is at a generic position.}
    \label{fig:figura}
\end{figure}

Exploiting rotational freedom, we fix the Earth at the origin, place pulsar $a$ on the $z$-axis, pulsar $b$ in the $xz$-plane, and pulsar $c$ at generic angles $(\chi_1,\chi_2)$,
see Fig.~\ref{fig:figura}:
\begin{equation}
    \mathbf{n}_{a} = (0,0,1), \quad 
    \mathbf{n}_{b} = (\sin\zeta, 0, \cos\zeta), \quad 
    \mathbf{n}_{c} = (\sin\chi_1 \cos\chi_2, \sin\chi_1 \sin\chi_2, \cos\chi_1).
\end{equation}
With this configuration, $\kappa^{\text{tens}}_{abc}=0$ after integration over the polarization angle $\psi$, consistent with the arguments of Sec.~\ref{sec_idea}. The scalar overlap function reduces to
\begin{eqnarray}
\label{res_kappta}
\kappa^{\text{scal}}_{abc}(\mathbf{n}_{a},\mathbf{n}_{b},\mathbf{n}_c)
=\tfrac{1}{12}\left(3+\mathbf{n}_a\cdot \mathbf{n}_b
+\mathbf{n}_a\cdot \mathbf{n}_c
+\mathbf{n}_b\cdot \mathbf{n}_c\right).
\end{eqnarray}
This remarkably simple expression generalizes in an intuitive way the two-point result of Eq.~\eqref{res_2ptpt}.  In fact, it maintains the monopolar structure
of a response function (in contrast with the typical quadrupolar ones found
in Einstein gravity).

\medskip
\noindent
{\bf Astrometry.}  
Astrometry provides a complementary probe of GW effects (see e.g.~\cite{Fakir:1993bj,Pyne:1995iy,Kaiser:1996wk,Jaffe:2004it,Book:2010pf,Shao:2014wja,Gaia:2016zol,Moore:2017ity,Klioner:2017asb,Theia:2017xtk,Qin:2018yhy,Garcia-Bellido:2021zgu,Malbet:2022lll,Darling:2018hmc,Wang:2022sxn,Wang:2020pmf,Caliskan:2023cqm,Pardo:2023cag,Jaraba:2023djs,Darling:2024myz,Aoyama:2021xhj,Mentasti:2023gmr,Inomata:2024kzr,Cruz:2024diu,Vaglio:2025tex}). Following the notation of~\cite{Book:2010pf}, the GW induces an apparent angular deflection
\begin{eqnarray}
    \delta n_i(\mathbf{n}_a,t)=\mathcal{R}_{ijk}(\mathbf{n}_a,\mathbf{n})\,h_{jk}(t,0),
\end{eqnarray}
in the direction of a star $\mathbf{n}_a$, where $\mathbf{n}$ is the GW propagation direction. The response tensor is
\begin{eqnarray}
    \mathcal{R}_{ijk}(\mathbf{n}_a, \mathbf{n}) = 
    \frac{n_{aj}}{2}\left[\frac{(n_{ai}+n_i)\,n_{ak}}{1+\mathbf{n}_a\cdot \mathbf{n}} - \delta_{ik}\right].
\end{eqnarray}
The equal-time three-point correlation of stellar deflections is then
\begin{eqnarray} \label{eq: Corr 3Star}
    \langle 
    \delta n_i(\mathbf{n}_a) \delta n_j(\mathbf{n}_b)
    \delta n_k(\mathbf{n}_c)\rangle
    =\mathcal{A}\,H_{ijk}(\mathbf{n}_a,\mathbf{n}_b,\mathbf{n}_c),
\end{eqnarray}
where $\mathcal{A}$ encodes integrals over the GW background, and 
\begin{eqnarray} \label{eq: 3Stars corr int}
    H_{ijk}(\mathbf{n}_a,\mathbf{n}_b,\mathbf{n}_c)=
    \int d^2\hat{n}\,\mathcal{K}_i(\mathbf{n}_a)\,
    \mathcal{K}_j(\mathbf{n}_b)\,
    \mathcal{K}_k(\mathbf{n}_c),
\end{eqnarray}
with
\begin{eqnarray} \label{def_ofK}
    \mathcal{K}_i(\mathbf{n}_a)=\mathcal{R}_{ijk}(\mathbf{n}_a)\,e^{\lambda}_{jk}.
\end{eqnarray}
with $\lambda$ the polarization index. We find that
performing the angular integral yields $H_{ijk}=0$ for all polarizations, scalars
included. Hence, astrometry alone possesses no non-trivial three-point overlap function and can instead serve as a null channel for noise characterization.

\smallskip
\noindent
{\bf Cross correlations between PTA and astrometry:}  
\begin{figure}[t!]
    \centering
    \includegraphics[width=0.5\linewidth]{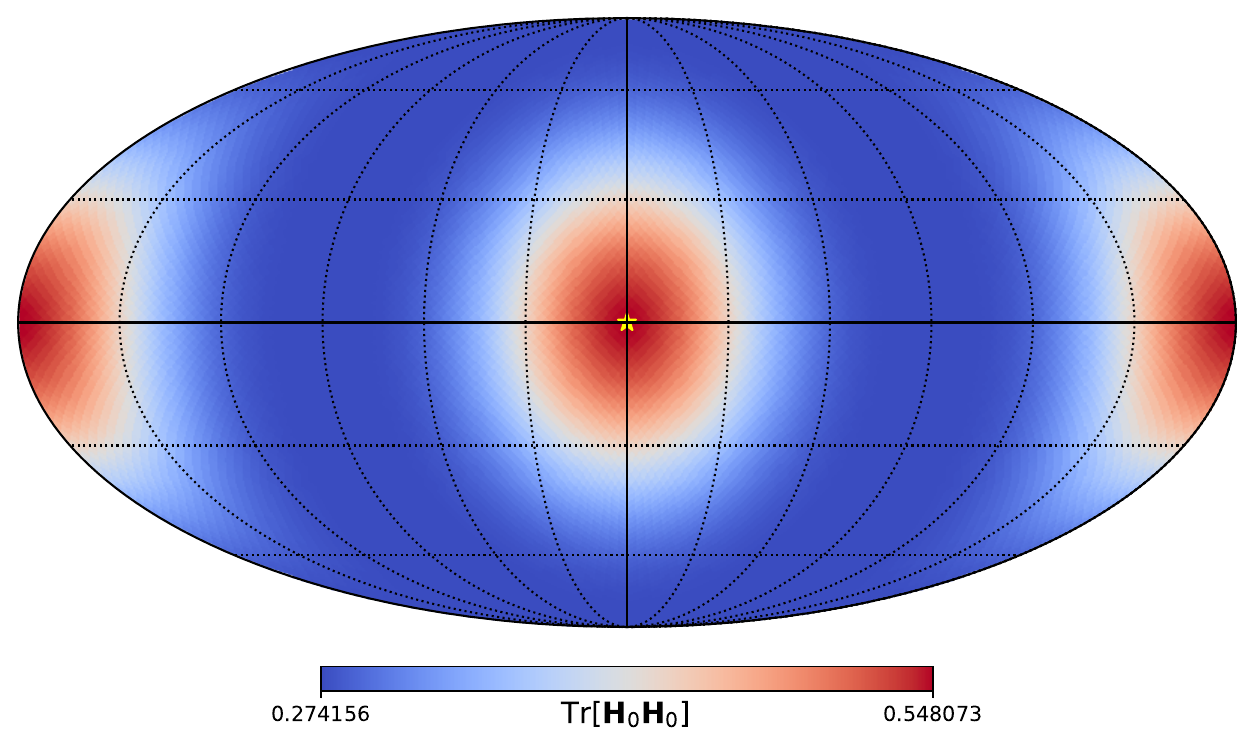}
    \caption{\small Mollweide
    projection of $\mathrm{Tr}[\mathbf{H}_0 \mathbf{H}_0]$ which correlate
    two stars with a pulsar position, see Eq.~\eqref{ov_ssp}. The result
    depends only on the star positions. For illustrative reasons, we fix one star  at the center
    of the plot, and we allow the direction of the second vary across the sky.}
    \label{fig:H0H0}
\end{figure}
The situation changes when pulsars and stars are cross correlated. In this case, three-point overlap functions for scalar polarizations are generically non-vanishing, and might be exploited to detect GW signals.  In fact, cross-correlations are especially
useful to calibrate measurements and reduce systematics. 

For the correlation of two stars and one pulsar, with $\mathbf{n}_{s_i}$ the direction of the $i$-th star and $\mathbf{n}_p$ the pulsar direction, we obtain
\begin{eqnarray}
    \langle \delta n(\mathbf{n}_{s_1},t)\,\delta n(\mathbf{n}_{s_2},t')\,z(\mathbf{n}_{p},f)\rangle
    = \mathcal{A}(f)\,H_{0}(\mathbf{n}_{s_1},\mathbf{n}_{s_2},\mathbf{n}_p),
\end{eqnarray}
with
\begin{eqnarray}
    \mathbf{H}_{0}(\mathbf{n}_{s_1},\mathbf{n}_{s_2},\mathbf{n}_p)
    =\int d^2\Omega\,\mathcal{K}_{i}(\mathbf{n}_{s_1})\,
    \mathcal{K}_{j}(\mathbf{n}_{s_2})\,
    F^{\text{b}}(\mathbf{n}_{p}),
\end{eqnarray}
where ${\cal K}_i$ is given in Eq.~\eqref{def_ofK} with $\lambda=b$, and $F^{\text{b}}$ is the PTA breathing-mode response defined in Eq.~\eqref{eq: PTA Corr Factor}. The angular integral evaluates to the following matrix components
\begin{eqnarray} \label{ov_ssp}
    H_{0\,ij}(\mathbf{n}_{s_1}, \mathbf{n}_{s_2})
    =\frac{\pi}{6}\Big[\delta_{ij}-n^{s_1}_i n^{s_1}_j
    -n^{s_2}_i n^{s_2}_j
    +(\mathbf{n}^{s_1}\cdot \mathbf{n}^{s_2})\,n^{s_1}_i n^{s_2}_j\Big],
\end{eqnarray}
which coincides with the two-point astrometry correlation of~\cite{Mihaylov_2018} and is independent of the PTA position. Curiously, the combination of scalar spin-0 polarization
tensors in the previous three-point expression manage to mimic the effect
of tensor spin-2 in the two-point reduction function. It would be interesting
to find a physical interpretation for this fact.
  Figure~\ref{fig:H0H0} illustrates $\mathrm{Tr}[\mathbf{H}_0 \mathbf{H}_0]$.  

\medskip

\begin{figure}[t!]
    \centering
    \includegraphics[width=0.45\linewidth]{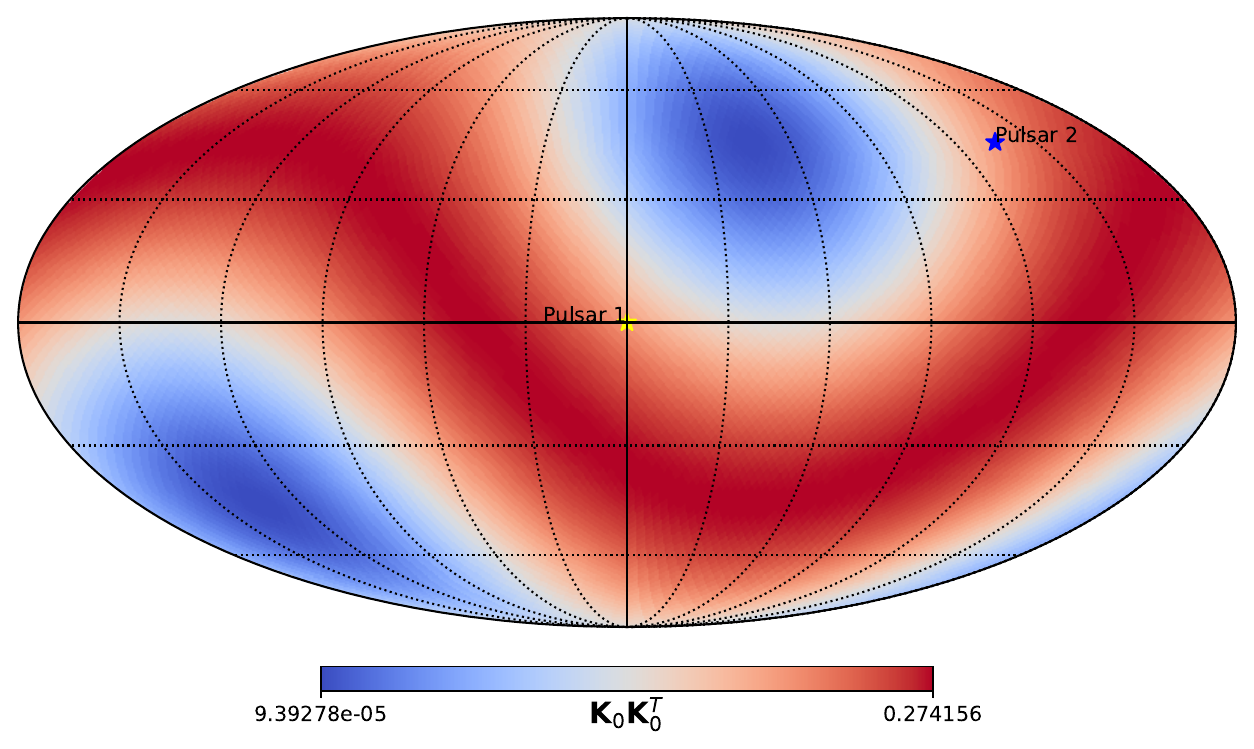}
    \includegraphics[width=0.45\linewidth]{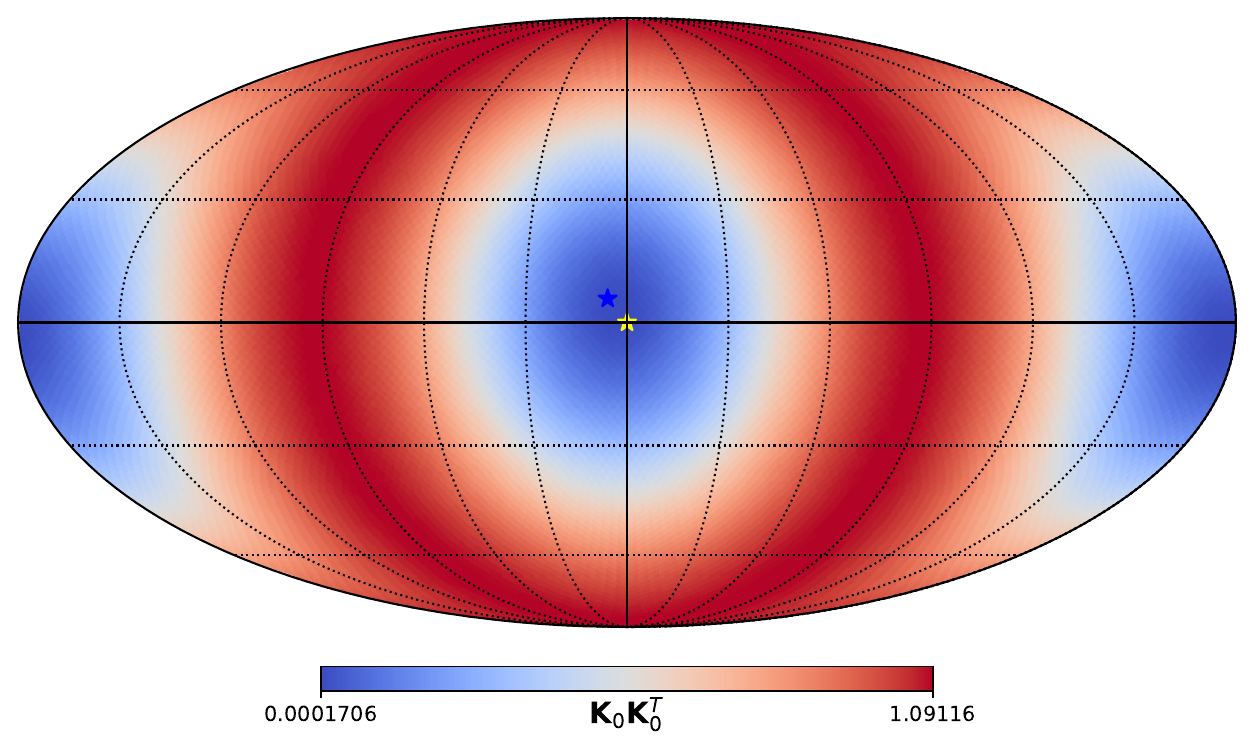}
    \caption{\small Mollweide
    projection of $\mathbf{K}_0 \cdot \mathbf{K}_0^T$ from Eq.~\eqref{def_ofK0}, for different pulsar--star configurations. In both panels, the first pulsar is fixed at the center. The left panel corresponds to the case where the second pulsar is far from the first, while the right panel shows the case where the two pulsars are close to each other. 
    Correspondingly, in both plots we vary the position
    of the star.}
    \label{fig:K0}
\end{figure}

Similarly, the correlation of two pulsars and one star is non-vanishing only for scalar polarizations. Restricting again to the breathing scalar mode, we find
\begin{eqnarray}
    \langle z(\mathbf{n}_{p_1})\,z(\mathbf{n}_{p_2})\,\delta n(\mathbf{n}_{s})\rangle
    = \mathcal{A}(f)\,\mathbf{K}_0(\mathbf{n}_{p_1},\mathbf{n}_{p_2},\mathbf{n}_s),
\end{eqnarray}
with the vector $ \mathbf{K}_0$ given by
\begin{eqnarray}
    \mathbf{K}_0(\mathbf{n}_{s},\mathbf{n}_{p_1},\mathbf{n}_{p_2})
    =\int d^2 \hat{n}\,\mathcal{ K}_i(\mathbf{n}_{s})\,
    F^{\text{b}}(\mathbf{n}_{p_1})\,F^{\text{b}}(\mathbf{n}_{p_2}),
\end{eqnarray}
which evaluates to
\begin{eqnarray}
    \mathbf{K}_0(\mathbf{n}_{s},\mathbf{n}_{p_1},\mathbf{n}_{p_2})
    =\frac{\pi}{6}\Big[(\mathbf{n}_{p_1}\times \mathbf{n}_{s})\times \mathbf{n}_{s}
    + (\mathbf{n}_{p_2}\times \mathbf{n}_{s})\times \mathbf{n}_{s}\Big].
    \label{def_ofK0}
\end{eqnarray}
Hence, again, a mathematically simple expression -- which
this time depends on the position of all three objects involved.  
The associated structure $\mathbf{K}_0 \cdot \mathbf{K}_0^T$ is shown in Fig.~\ref{fig:K0}.  

\smallskip
\noindent
{\bf Summary.}  
In summary, we have obtained the three-point overlap reduction functions for PTA and astrometry.  Overlap functions vanish for tensor and vector polarizations but are generically non-vanishing for scalar polarizations. 
We discussed
their geometrical interpretation, if any, in terms
of properties of scalar polarization
tensors.  We point out that mixed correlators (tensor--tensor--scalar or vector--vector--scalar) can also be non-zero, but a scalar mode must always be present. Thus, any non-vanishing detection of a three-point correlation function would provide a clear and distinctive signature of scalar GW polarizations.



\section{Detecting  a gravitational wave three-point function}
\label{sec_opt}

In the previous section we discussed the response of gravitational-wave detectors to the GW three-point function - mostly focussing on pulsar timing arrays and astrometry -- and pointed out their potential use as a smoking gun for scalar polarizations. We now examine a convenient estimator for this observable.  We correspondingly design 
a simple likelihood function, and we develop preliminary Fisher forecasts 
in an idealized sitation, so to explore
 the detectability of GW three-point function with  pulsar timing array measurements.

\subsection{Building an optimal estimator}
\label{sec_buoe}

We  construct an optimal estimator for the GW three-point function, following the approach of \cite{Powell:2019kid}, but adapting it to the case of scalar modes. For building such estimator, in this section we keep our discussion general,  and we do not
need to specify  the GW detector we focus on. 
Let  
\begin{equation}
\Sigma(t) = s(t) + n(t)
\end{equation}
denote the time-domain output of a single GW detector, where $s(t)$ is the gravitational-wave signal and $n(t)$ represents the instrumental noise. We consider the cross-correlation of {\it three} such measurements $\Sigma(t)$, assuming that the noise $n(t)$ is stationary, Gaussian, and uncorrelated between different detectors
~\footnote{ A Gaussian noise hypothesis allows us -- by measuring signal
three-point functions --  to avoid
systematic uncertainties  on pulsar timing and solar system
ephemeris which can  affect measurements of two-point correlation functions. 
 We point out though  that recent
interesting analysis are taking into account
more general possibilities for noise statistics -- see e.g. \cite{Ghonge:2023ksb,Karnesis:2024pxh,Xue:2024qtx,Bernardo:2024uiq,Falxa:2025qxr}. }. We focus on the noise-dominated regime, in which the variance is set by the noise, while the expectation value of the statistic is determined entirely by the signal $s(t)$.  Signals from different detectors may exhibit non-Gaussian correlations
with non-vanishing three-point functions $\langle s^3 \rangle$, assumed to be stationary as in the hypothesis  outlined in Section~\ref{sec_3ORF}.

\smallskip

A combination of three copies of  $\Sigma(t) $ provides
the estimator we are interested to study, which we use to test
the presence of scalar polarizations in GW. We build the quantity
\begin{equation}
S_{abc} = \int_{-T/2}^{T/2} dt_1 \int_{-T/2}^{T/2} dt_2 \int_{-T/2}^{T/2} dt_3 \, 
\Sigma_a(t_1) \, \Sigma_b(t_2) \, \Sigma_c(t_3) \, Q(t_2 - t_1, t_3 - t_1) ,
\label{def_sabc}
\end{equation}
where $T$
is the time duration of our  measurement, and  $Q$ is a filter function -- to be determined -- depending only on the time differences given we work on the hypothesis of stationarity. It vanishes  for large separations $|t_i - t_j|$. The goal is to choose the filter function $Q$ to maximize the response to the signal. 
  For the moment, we focus on a single
triplet of measurements, $(abc)$ with three different GW detectors.

We Fourier transform the time series as
\begin{equation}
\Sigma(t) = \int df \, e^{2 \pi i f t} \, \tilde{\Sigma}(f) ,
\end{equation}
which yields
\begin{equation}
S_{abc} = \int df_1 \, df_2 \, df_3 \,
\delta_T(f_1 + f_2 + f_3) \,
\tilde{\Sigma}_a(f_1) \, \tilde{\Sigma}_b(f_2) \, \tilde{\Sigma}_c(f_3) \,
\tilde{Q}^*(f_2, f_3) ,
\end{equation}
where we have introduced the finite-time delta function (and $T$
the measurement duration)
\begin{equation}
\delta_T(f) \equiv \int_{-T/2}^{T/2} dt \, e^{2 \pi i f t}, 
\qquad \delta_T(0) = T .
\end{equation}
In the noise-dominated regime, the mean and variance of $S_{abc}$ are
\begin{align}
\mu &= \langle S_{abc} \rangle, \\
\label{eq_defva}
\sigma^2 &= \langle S_{abc}^2 \rangle -\mu^2 .
\end{align}
Since the noise is Gaussian, $\mu$ is determined solely by the signal, while $\sigma^2$ is determined by the noise (which we assume much larger
in amplitude than the signal -- hence we can neglect the contribution
proportional to $\mu^2$ in Eq.~\eqref{eq_defva}).  The signal-to-noise ratio (SNR) reads
\begin{equation}
\mathrm{SNR} = \frac{\mu}{\sigma} ,
\end{equation}
and constitutes the quantity to maximise by appropriately choosing the filter
function.

A direct calculation gives
\begin{equation}
\mu = T \, \kappa_{abc} \, \int df_1 \, df_2 \, B_{b}(f_1, f_2, f_3) \, \tilde{Q}^*(f_1, f_2) ,
\end{equation}
where $\kappa_{abc}$ is the three-detector overlap reduction function and $ B_{b}(f_1, f_2, f_3)$ is the  breathing-mode bispectrum defined in Eq.~\eqref{def_bis},
satisfying~\footnote{Since $f_3= -f_1-f_2$, in what follows
we understand the dependence of $f_3$ on $B_{b}$.} the condition $f_1+f_2+f_3=0$.  
Indicating with $\sigma^2_a(f_1)$ the noise spectrum,
the noise two-point function is defined as
\begin{equation}
\langle n_a(f_1) n_b(f_2) \rangle = \delta(f_1 + f_2) \, \delta_{ab} \, \sigma^2_a(f_1) \,,
\end{equation} 
leading to the total  variance (recall that we work under the hypothesis that the noise has Gaussian distribution) 
\begin{equation}
\sigma^2 = T \int df_1 \, df_2 \, N_{abc}(f_1, f_2) \, |Q(f_1, f_2)|^2 ,
\end{equation}
where
\begin{equation}
N_{abc}(f_1, f_2) \equiv \sigma^2_a(f_1) \, \sigma^2_b(f_2) \, \sigma^2_c(f_1 + f_2) + \mathrm{perms} .
\end{equation}
Thus, the SNR becomes
\begin{equation}
\mathrm{SNR} =
\sqrt{T} \,
\frac{\kappa_{abc} \, \int df_1 \, df_2 \, B_{b}(f_1, f_2) \, \tilde{Q}^*(f_1, f_2)}
{\left[ \int df_1 \, df_2 \, N_{abc}(f_1, f_2) \, |Q(f_1, f_2)|^2 \right]^{1/2}} .
\end{equation}

Adopting  the Wiener filtering technique (see e.g. \cite{Allen:1997ad} for a clear discussion in a similar context), we introduce the positive-definite inner product
\begin{equation}
(C, D) \equiv \int df_1 \, df_2 \, C(f_1, f_2) \, D^*(f_1, f_2) \, N_{abc}(f_1, f_2) ,
\end{equation}
so that
\begin{equation}
\mathrm{SNR} = \sqrt{T} \,
\frac{\left( \kappa_{abc} \, B_{b} / N_{abc} , Q \right)}
{\left[ \left( Q , Q \right) \right]^{1/2}} .
\end{equation}
The optimal filter we are searching for  is then
\begin{equation}
\label{eq_opf}
Q(f_1, f_2) = \frac{\kappa_{abc} \, B_{b}(f_1, f_2)}{N_{abc}(f_1, f_2)} ,
\end{equation}
which yields the maximum achievable SNR in our context:
\begin{equation}
\mathrm{SNR}_{\mathrm{max}} = \sqrt{T} \,
\left[ \int df_1 \, df_2 \, \frac{\left( \kappa_{abc} \, B_{b}(f_1, f_2) \right)^2}{N_{abc}(f_1, f_2)} \right]^{1/2} \,.
\end{equation}

A further simplification occurs for frequency-independent noise,
$\langle n_a(f_1) n_b(f_2) \rangle = \delta(f_1 + f_2) \, \delta_{ab} \, \sigma_a^2$,
in which case
\begin{equation}
N_{abc}(f_1, f_2) = 3 \, \sigma_a^2 \, \sigma_b^2 \, \sigma_c^2 .
\end{equation}
and 
\begin{equation}
\mathrm{SNR}_{\mathrm{max}} = \sqrt{\frac{T}{3}} \,
\left[ \int df_1 \, df_2 \, \frac{\left( \kappa_{abc} \, B_{b}(f_1, f_2) \right)^2}{ \sigma_a^2 \, \sigma_b^2 \, \sigma_c^2} \right]^{1/2} \,.
\label{eq_opesi}
\end{equation}
for measurements from a single triplet of detectors. If multiple GW detector triplets are in principle available -- as in the case of PTA --  we can  increase 
 the total SNR by summing over all possible combinations. 

In summary, we have constructed an optimal estimator for detecting the scalar bispectrum $ B_{b}$, 
which serves as a direct diagnostic of scalar polarizations in the SGWB signal. 
The procedure consists of forming a suitable combination of three measurements $\Sigma(t)$ 
and integrating over time, see  Eq.~\eqref{def_sabc}. 
After Fourier transforming and applying the optimal filter defined in Eq.~\eqref{eq_opf}, 
we obtain the best possible measurement of $ B_{b}$.  On this basis, we next develop  a simple forecast  for the detectability of scalar polarizations 
with idealized future PTA measurements, employing the Fisher formalism.

\subsection{Fisher forecasts for detecting scalar polarizations with
 PTA}
\label{sec_like}

Based on the results we derived above, and under a set of simplifying assumptions
we are going to describe, we  construct a likelihood for our estimator of scalar polarizations in GW experiments. Such likelihood might serve as the basis for deriving idealized Fisher forecasts for the detectability of scalar modes with PTA by taking
signal three-point functions. 

We focus on measuring the scalar bispectrum $B_{b}$, see  Eq.~\eqref{def_bis}, as a diagnostic of the presence of scalar polarizations.  For simplicity, in this section we assume 
a `power-law' behaviour for this quantity  \footnote{In Sec.~\ref{sec_cosex} we discuss a more realistic cosmological setup leading to a richer structure for the scalar-polarization bispectrum. Our analysis  can be straightforwardly extended to that case,
as well as to other theoretically motivated  setup.}, with an Ansatz
\be
B_{b}(f_1, f_2)\,=\,\frac{P_0^3}{f_1^{3-n_1} \,f_2^{3-n_2}}
\ee
parametrized by two spectral indexes  $n_{1,2}$ in Eq.~\eqref{def_kaB}. 
We indicate the bispectrum amplitude as cube $P_0^3$
to indicate it originates  from a three point function -- 
an explicit example is developed in Section~\ref{sec_cosex}.

We would like to quantify the bispectrum 
amplitude $P_0$ which can in principle be measured with PTA data, under the
simple Ansatz above. 
We assume that the likelihood for $B_{b}$ follows a Gaussian distribution, with variance determined by the experimental noise properties described earlier. 
Hence
 our approach extends that of \cite{Anholm:2008wy,Ali-Haimoud:2020ozu,Ali-Haimoud:2020iyz,Cruz:2024svc,Cruz:2024esk}, who considered Gaussian likelihoods for the intensity and polarization of the SGWB in the context of PTAs. We write the logarithm of the likelihood ${\cal L}$ as
\bea \label{eq:likelihood_P0}
-2 \,\ln {\cal L}
&=&{\rm const.}+
\sum_{f_1, f_2}\,\sum_{AB}
\left({\cal R}_A-\kappa_A\cdot  B_{s}
\right)\,C_{AB}^{-1}\,
\left({\cal R}_B-\kappa_B\cdot  B_{s}
\right) ,
\eea
where
${\cal R}_A$ are the Fourier transform of time-integrated three point functions of GW measurements, and the sums are over pulsar triplets, denoted by $A=(abc)$.
The dot denotes the  integrated quantity
\be
\label{def_kaB}
\kappa_A\cdot  B_{s}
\,=\,P_0^3\,\kappa_{A}\,\int_{f_1-\Delta f/2}^{f_1+\Delta f/2}
\int_{f_2-\Delta f/2}^{f_2+\Delta f/2}
\,\frac{d \tilde f_1 d \tilde f_2}{\tilde f_1^{3-n_1} \,\tilde f_2^{3-n_2}} ,
\ee
generalizing \cite{Cruz:2024svc,Cruz:2024esk} to the case of the bispectrum.  
Each integration runs over a small frequency interval $\Delta f$, and the sum in Eq.~\eqref{eq:likelihood_P0} covers all such intervals. 
The quantity $\kappa_A$ denotes the PTA response to scalar polarizations in three-point measurements, as in Eq.~\eqref{res_kappta}.

The inverse covariance matrix is  deduced from the results of Sec.~\ref{sec_buoe}. Assuming, for simplicity, that all pulsars are monitored over the same observation time $T$, we have
\be
\label{def_icm}
C_{AB}^{-1}\,=\,
\frac{2 T \,\Delta f^2}{3\,{\bf R}_A^N
{\bf R}_B^N
}\,\delta_{AB} ,
\ee
where $A,B$ denote pulsar triplets. This quantity also depends on the frequency interval $\Delta f$ used to bin the frequency bands, over which
we sum. The factors ${\bf R}_A^{N}$ in the previous equation encode the intrinsic pulsar noise  and in the weak-signal limit we obtain
\be
{\bf R}_A^N\simeq
\left(\sigma_a^2 \sigma_b^2 \sigma_c^2 \right)^{1/2} ,
\ee
with $\sigma_a$ the band-integrated noise variance of pulsar $a$. 

\begin{figure}[t!]
    \centering
        \includegraphics[width=0.4\linewidth]{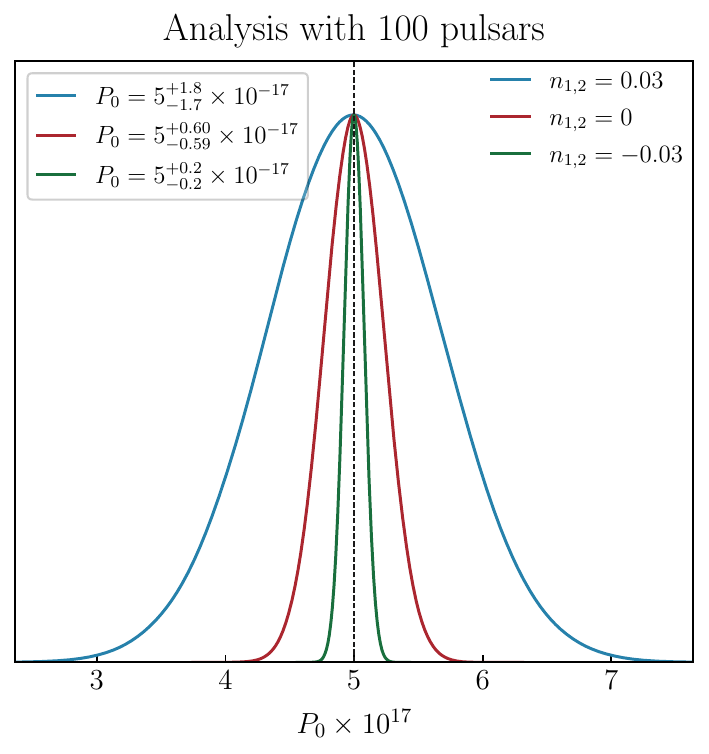}
        \includegraphics[width=0.4\linewidth]{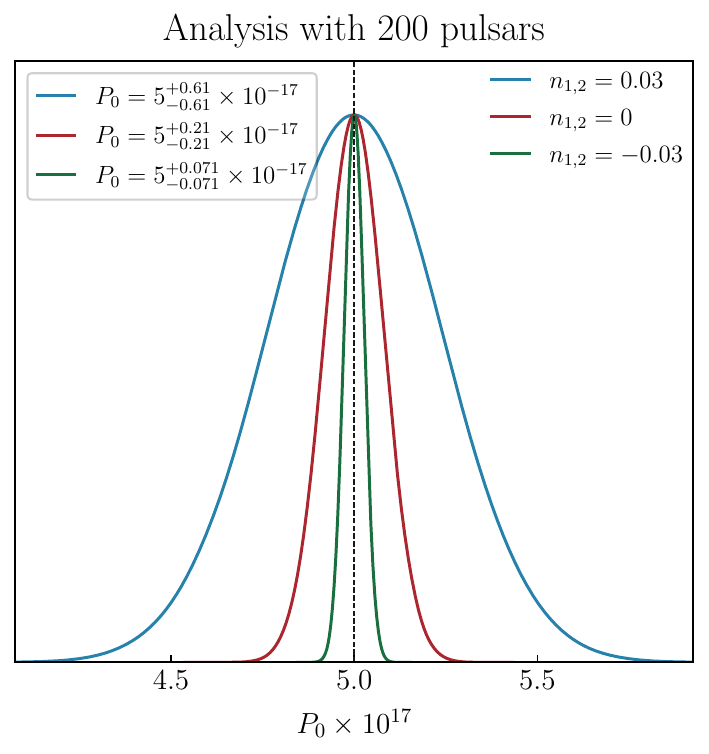}
   \caption{\small Constraints on the parameter $P_0$ of Eq.~\eqref{def_kaB} at  3-$\sigma$ confidence level from the Fisher forecast described in Sec.~\ref{sec_like}. In the two plots we represent results for two pulsar populations and different choices of spectral indices. We take $n_1=n_2$ in the cases shown. The confidence intervals are displayed in the left box of each panel. In all cases, the fiducial value is fixed to $P_0 = 5 \times 10^{-17}$.}
    \label{fig:P0_f1f2}
\end{figure}

Starting from the likelihood \eqref{eq:likelihood_P0}, we compute the Fisher matrix associated with the parameters of interest. In this initial study, we focus on the single parameter, the amplitude $P_0$ of the  bispectrum with structure as in Eq.~\eqref{def_kaB}.
We keep the spectral indexes $n_{1,2}$ fixed, and we estimate
the sensitivity of measurements to $P_0$. 
 The (single) Fisher matrix component for each frequency bin $\Delta f$ is summed over all bins and all pulsar triplets:
\be
    \fc_{P_{0} P_{0}} = \left\langle -\frac{\partial^2 \ln {\cal L} }{\partial P_0^2} \right \rangle\, = \sum_{f_1, f_2}\,\sum_{AB} \frac{ 2 T  \Delta f^2}{{\bf R}_A^N
{\bf R}_B^N} \left(\frac{6 P_0^4\, \kappa_A \kappa_B\,\delta_{AB}}{ f_1^{6- 2 n_1} f_2^{6- 2 n_2}}\right).
\ee

For the forecast, we adopt a frequency binning of $\Delta f = 1/T$, with a common observation time of $T = 15 \, \text{years}$. As representative experimental
setup, we take the noise amplitude associated with  pulsar J1012-4235 from the NG15 results~\cite{NANOGrav:2023ctt} as the common noise model for all pulsars. For simplicity we assume pulsars are randomly distributed on the sky, and we consider two ensembles of $100$ and $200$ pulsars as representative
of forthcoming PTA experiments. We find that the results are very  sensitive to the spectral indices $n_{1,2}$, indicating that the frequency dependence of the bispectrum plays an important role in determining its observability. Hence, the results of measurements
will depend on the theoretical models one considers.  
The forecasts are summarized in Fig.~\ref{fig:P0_f1f2}
obtained using the \texttt{GetDist} package \cite{Lewis:2019xzd}, where we learn that values of $P_0$ of order $5\times 10^{-17}$ are in principle detectable, with variations in the error bar size of order unity depending on the number of pulsars and on the spectral indices $n_{1}, n_{2}$ in Eq.~\eqref{def_kaB}.  

Although this simple Fisher analysis can be improved in several directions, it provides a concrete illustration that, in principle, measuring the GW three-point function can serve as a useful diagnostic of scalar polarizations, provided the bispectrum amplitude is sufficiently large  to be detectable.


\section{A cosmological  source for  three-point functions}
\label{sec_cosex}
After outlining in general terms how measurements of the GW three-point function can reveal the presence of scalar polarizations, in this section we develop an explicit cosmological example that can generate an enhanced three-point function for scalar GW modes.  Such setup
might be used as benchmark cosmological scenario to search
for additional polarizations in the SGWB
implementing our method. 
Cosmological sources of the SGWB (see \cite{Caprini:2018mtu} for a comprehensive review) may account for at least part  of the PTA-detected SGWB signal -- see e.g. the early studies \cite{Ellis:2023oxs,Figueroa:2023zhu}.  
If this were the case, it may be of course  unfeasible to apply methods that rely on assumptions about black hole merger waveforms (see, e.g., the recent \cite{Liang:2024sfn}
and references therein) in order to extract information on additional polarizations.  
The possibility of cosmological
sources contributing to the SGWB then further motivates the new approach introduced in this work.

\subsection{Second-order perturbations and gravitational waves}

We expect that any non-linear source of a stochastic gravitational-wave background (SGWB) 
 produces non-Gaussianities in the signal. 
If detectable, such non-Gaussian features can yield valuable insights into the physical origin of the SGWB, 
as well as important information about the underlying theory of gravitational interactions. 
Indeed, a variety of theoretical studies have quantified GW non-Gaussianity from both cosmological and astrophysical sources, 
and explored its possible observational signatures --  see Section~\ref{sec_revng}.

\smallskip
In this work we investigate, for the first time, tensor non-Gaussianities in a cosmological setup based on GW induced at second order  fluctuations, a framework 
which leads to  a particularly transparent and instructive calculation.
Our aim here is twofold:
\begin{enumerate}
\item Apply the theoretical approach of second order induced GW to the case of extra
GW polarizations, with the aim of  testing the existence of additional polarizations in GW measurements
through the methods discussed in the previous sections~\footnote{See also the study
\cite{Kugarajh:2025rbt} which explores scalar-induced
gravitational waves in alternative theories of gravity.}.
\item Clarify the importance  of  stationary three-point functions for obtaining a measurable signal, explicitly addressing in this context  arguments
first  developed in \cite{Bartolo:2018rku}.
\end{enumerate}
If gravitational waves carry extra polarizations—beyond the spin-2 modes of General Relativity— 
they are  sourced at second order by fluctuations of a Gaussian  field in the early Universe, 
in complete analogy with the standard generation of spin-2 polarizations. Instead of the more commonly-considered
scalar perturbations,
we focus on a spectator vector source, which
will turn out to be easier to handle
for our purposes. 
Schematically, the GW tensor perturbation $h_{ij}$ is sourced by an operator quadratic in a vector field $v_i$: 
\begin{equation}
\label{eq_scA2}
h_{ij} \,\sim\, v_i v_j \,,
\end{equation}
so that the GW two- and three-point functions are proportional to the vector four- and six-point functions, respectively. Applying Wick’s theorem, a $2n$-point function of a Gaussian vector field decomposes into products of $n$ two-point functions. Besides the case $n=2$ (yielding the GW power spectrum),  for our purposes we also examine the bispectrum case with $n=3$. 

\smallskip

\noindent
{\bf Primordial magnetic fields as GW source.}
We illustrate this mechanism in a context  motivated by primordial magnetogenesis. This framework postulates that cosmological magnetic fields—capable of explaining the observed large-scale magnetic fields in the Universe—were generated during cosmic inflation. See, e.g.,~\cite{Durrer:2013pga} for a review. We assume that some early-Universe process produces a large-scale magnetic field ${\bf B}$, which in turn sources the physical GW polarizations at second order in perturbations. To compute the resulting GW signal, we adapt the methods already developed  for scalar-induced GW scenarios, where enhanced curvature perturbations source GWs after inflation. This line of research has a long history, see~\cite{Domenech:2021ztg} for a review,
and \cite{LISACosmologyWorkingGroup:2025vdz,Ghaleb:2025xqn} for recent techniques for reconstructing
the signal. Here we apply the formalism to the non-adiabatic case of magnetic-field sources, as studied in several earlier works~\cite{Durrer:1999bk, Caprini:2001nb,Mack:2001gc,  Pogosian:2001np, Caprini:2003vc, Shaw:2009nf, Saga:2018ont,Bhaumik:2025kuj, Maiti:2025cbi}, and initiate the calculation of the GW bispectrum for the scalar (breathing) polarization in this context.  We expand
the breathing mode contribution to the GW in Fourier space as~\footnote{In this section, in order to align with the literature on field-theoretic treatments of second-order GW sources, we perform a Fourier transform over the three spatial dimensions, rather than the four-dimensional transform employed in Section~\ref{sec_3ORF}.}:
\begin{align}
\label{eq_fmd}
h_{ij}(\tau, {\bf x}) &= \int \frac{d^3{\bf k}}{(2\pi)^{3/2}} \, e^{i {\bf k} \cdot {\bf x}} \,
\mathbf{e}^{(b)}_{ij}({\bf k}) \, {h}_{\bf k}(\tau) \,,
\end{align}
where we retain only the scalar breathing mode, which is the focus of our analysis since only the presence of scalar polarization can lead to non-vanishing
detector response functions 
in the case of three point functions (recall Section~\ref{sec_3ORF}).

Since the magnetic field is governed by the Maxwell action in curved spacetime—quadratic in the vector fields—we assume that it obeys Gaussian statistics, with a two-point function satisfying the relation  \cite{Mack:2001gc}
\begin{align}
\label{eq_mfa}
\langle B_i({\bf k}) B_j({\bf q}) \rangle'_{\bf k+q=0} &= P_{0}\,\pi_{ij}({\bf k}) \, f(k) \,,
\end{align}
where the prime indicates that the momentum-conserving delta function has been omitted. We place as overall factor a constant $ P_{0}$  -- which
characterizes the magnetic spectrum amplitude at large scales -- and we denote
as $\pi_{ij}$  the symmetric tensor
\begin{align}
\pi_{ij}({\bf k}) &= \delta_{ij} - \frac{k_i k_j}{k^2} \,,
\end{align}
with $k = \sqrt{{\bf k} \cdot {\bf k}}$. We indicate with  $f(k)$ a (model-dependent)
dimensionless function controlling the scale dependence of the magnetic
spectrum,  normalizing it by imposing it has unit value
at large, cosmic microwave background scales:  $f(k_{\rm CMB})\,=\,1$. While often a simple power-law scaling is assumed
for $f(k)$, one can also consider
richer scenarios, as e.g. \cite{Atkins:2025pvg}.

\smallskip
The magnetic field acts as a source term for the scalar polarization in the GW evolution equation in Fourier space (recall the Fourier  mode definition of Eq.~\eqref{eq_fmd}):
\begin{align}
\label{eq_eveqh}
h_{\bf k}'' + 2{\cal H}\,h_{\bf k}' + k^2\,h_{\bf k} 
&= \frac{S_{\bf k}}{a^2(\tau)} \,,
\end{align}
where ${S}_{\bf k} = {e}^{(b)}_{ij} \, \tau^{(B)}_{ij}$, and the magnetic-field energy–momentum tensor is (see e.g. \cite{Mack:2001gc})
\begin{align}
\label{eq_emtmf}
\tau_{ij}^{(B)}({\bf k}) &= \frac{1}{4\pi} \int \frac{d^3 {\bf p}}{(2\pi)^3} 
\left[ B_i({\bf p}) B_j({\bf k}-{\bf p}) - \frac{\delta_{ij}}{2} B_m({\bf p}) B_m({\bf k}-{\bf p}) \right] .
\end{align}
Notice that the scalar
polarization tensor
\(
\mathbf{e}^{(b)}_{ij}(\hat{k}) = \pi_{ij} (\hat{k})  
\)
projects the magnetic field energy moment tensor along the components
of the breathing scalar mode.
The source term becomes 
\begin{align}
\label{eq_gwsor}
{S}_{\bf k}(\tau) &= -\frac{1}{4\pi\,k^2} \int \frac{d^3 {\bf p}}{(2\pi)^3} 
\left[ {\bf k} \cdot {\bf B}({\bf p}) \right] \left[ {\bf k} \cdot {\bf B}({\bf k} - {\bf p}) \right] .
\end{align}
Equation~\eqref{eq_eveqh} can be solved formally as
\begin{align}
\label{eq_frh}
{h}_{\bf k}(\tau) &= \frac{1}{a(\tau)} \int_{\tau_R}^\tau d\tau' \, \frac{g_k(\tau, \tau')}{a(\tau')} \, {S}_{\bf k}(\tau') \,,
\end{align}
where $g_k$ is the Green’s function for the system under consideration. In what follows, we focus on radiation domination, and $\tau_R$ denotes the time of instantaneous reheating
at the end of inflation.  Equation \eqref{eq_frh} makes precise the schematic relation~\eqref{eq_scA2}, showing that the GW signal is indeed sourced by the quadratic combination of vector modes. During radiation domination,
\begin{align}
g_k(\tau, \tau') &= \frac{1}{k} \left[ \sin(k\tau) \cos(k\tau') - \sin(k\tau') \cos(k\tau) \right] .
\end{align}
 On this basis, we now proceed to compute explicitly the two- and three-point functions of the GW solution in Eq.~\eqref{eq_frh}. We expect that these correlators depend on the square
 and the cube of the magnetic vector source, hence -- according
 to Eq. \eqref{eq_mfa} -- they are proportional to $P_0^2$ and
 $P_0^3$ respectively.

 \subsection{The gravitational wave two-point function}
 \label{sec_GWtpf}
The GW two-point function is a key observable, as it is directly related to the GW energy
density. Several works~\cite{Durrer:1999bk, Caprini:2001nb,Mack:2001gc,  Pogosian:2001np, Caprini:2003vc, Shaw:2009nf, Saga:2018ont,Bhaumik:2025kuj, Maiti:2025cbi} have investigated how primordial magnetic fields can act as a source for this quantity. 
As a warm-up, we extend here the discussion of~\cite{Atkins:2025pvg} to the computation of two-point correlators
involving scalar polarizations, as induced by primordial magnetic fields. 
This case provides a useful starting point, since it will be straightforwardly generalized to the three-point function in Section~\ref{sec_tpfgw}. 

We write the GW scalar polarization power spectrum as
\begin{align}
P_{ h}(k)\,\equiv\,\langle
 h_{\bf k}(\tau) 
 h_{\bf q}(\tau) 
\rangle'_{\bf k=-\bf q}
&\equiv
\frac{1}{a^2(\tau)}
\int d \tau_1 \, d \tau_2 \,
\frac{g_{ k}(\tau, \tau_1)}{a(\tau_1)}
\frac{g_{ q}(\tau, \tau_2)}{a(\tau_2)} 
\,
\langle
{S}_{\bf k}
{S}_{\bf q}
\rangle'_{\bf k=-\bf q} 
\nonumber \\
&= \,\frac{1}{a^2(\tau)}\,{\cal I}_{(2)}(\tau)\;
\langle
{S}_{\bf k}
{S}_{\bf q}
\rangle'_{\bf k=-\bf q} \,,
\label{eq_ts2s}
\end{align}
with
\begin{equation}
\label{def_tim2pt}
{\cal I}_{(2)}(k,\tau)
= \left(  \int_{\tau_R}^\tau d\tau_1 \, \frac{g_{ k}(\tau, \tau_1)}{a(\tau_1)} \right)^2 .
\end{equation}
We evaluate the result at late times, $\tau/|\tau_R|\gg1$, during radiation domination. 
These expressions are structurally  similar to the formulas obtained for spin-2 polarizations, differing only by overall coefficients arising from distinct polarization tensors.  

Equation~\eqref{eq_ts2s} shows that the time and momentum integrals factorize: all the time dependence is contained in ${\cal I}_{(2)}$, while the momentum dependence resides in the source two-point function $\langle
{S}_{\bf k}
{S}_{\bf q}
\rangle'_{\bf k+\bf q=0}$. 
This separation is especially convenient, since the two quantities can be computed independently -- a property which will be especially useful in  Section~\ref{sec_tpfgw} when discussing the detectability of the signal three-point correlators.

\smallskip
\noindent
{\bf The time integral.}
We begin with  to the time integral in Eq.~\eqref{def_tim2pt}, evaluated in the limit of large conformal time $\tau$, well within the radiation-dominated era. 
This expression produces contributions involving $\sin(k\tau)$, $\cos(k\tau)$, and their squares. 
Terms linear in these oscillatory functions vanish upon averaging over rapid oscillations, whereas quadratic terms average to $1/2$. 
The result is
\begin{align}
\label{eq_logd1}
{\cal I}_{(2)} (k,\tau)
&=\frac{1}{2\,k^2\,a^4 H^2}
 \left[ \text{Ci}(-k {\tau_R})^2+\left( \frac{\pi}{2} -\text{Si}(-k {\tau_R})\right)^2\right],
\end{align}
where $\text{Ci}(x)$ and $\text{Si}(x)$ denote the cosine and sine integral functions, respectively.  
This expression exhibits a logarithmic divergence for small $|k \tau_R|$, since $\text{Ci}(x)\sim\ln(x)$ as $x \to 0$.

\smallskip
\noindent
{\bf Contributions from the source.}
We now turn to the source two-point function, which takes the form of a convolution integral:
\begin{align}
\langle
\bar{S}_{\bf k}
\bar{S}_{\bf q}
\rangle'_{\bf k+\bf q=0}
&=
\frac{k_{a}k_{b}k_{c}k_{d}}{(4 \pi)^2\,k^4}
\int \frac{d^3 \bf p_1}{(2 \pi)^3}
\int \frac{d^3 \bf p_2}{(2 \pi)^3}
\left\langle 
B_a(\mathbf{p}_1)\,
B_b(\mathbf{k}-\mathbf{p}_1)\,
B_c(\mathbf{p}_2)\,
B_d(\mathbf{k}+\mathbf{p}_2)
\right\rangle
\\
&=\frac{ P_{0}^2}{8 \pi^2}
F(k)
\end{align}
with
\be
F(k)\equiv\int\!\frac{d^3{\bf p}_1}{(2\pi)^3}\;
{f({p}_1)\,f(|\mathbf{k}-\mathbf{p}_1|)}
\,A\big(\mathbf{k},-\mathbf{k};\mathbf{p}_1\big)\;
A\big(\mathbf{k},-\mathbf{k};\mathbf{k}-\mathbf{p}_1\big) ,
\ee
where
\be
\label{def_fA}
A(\mathbf{k}_1,\mathbf{k}_2;\mathbf{u})
=\frac{\mathbf{k}_1\!\cdot\!\mathbf{k}_2}{k_1 k_2}
    -\frac{(\mathbf{k}_1\!\cdot\!\mathbf{u})\,(\mathbf{k}_2\!\cdot\!\mathbf{u})}{u^2\,k_1 k_2} \,.
\ee
Hence, the scale dependence of the source correlator is set by the magnetic-field power spectrum $P_B(k)$ proportional to the function $f(k)$ (recall
Eq.\eqref{eq_mfa}).

\smallskip
\smallskip
\noindent
{\bf The energy density.}
Collecting these results, and retaining only the logarithmically enhanced contribution from Eq.~\eqref{eq_logd1}, the GW  scalar-polarization power spectrum can be written as
\bea
P_{ h}\,=\,\left(\frac{a H}{k} \right)^2 \frac{P_{0}^2 }{8 \pi} 
\frac{ \ln^2\left(k^2 {\tau^2_R} \right)}{a^8\,H^4}\,F(k)\,.
\eea
This representation is particularly convenient for expressing all results in terms of energy densities, defined for a species $A$ as
\be
\Omega_{A}\,=\,\frac{1}{\rho_{\rm cr}}\,\frac{d \rho_A}{d \,\ln f}\,.
\ee
The
energy density in the scalar polarization is related to its power spectrum through (we use the conventions of~\cite{Kohri:2018awv})
\be
 \Omega_{\rm GW}^{(b)}\,=\,\frac{1}{24} \left(\frac{k}{a H} \right)^2\,P_{ h}\,.
\ee
(The overall factor scaling as $k^2/(a H)^2$  is due to the fact that the GW energy 
depends on (conformal) time derivatives of the GW mode $h_{ij}$ squared.)
The
magnetic-field energy density on large scales is given by 
$\Omega_B= P_{0}/(3 H_0)\,$ (see e.g. \cite{Caprini:2001nb}). 
Parametrizing the scale factor during radiation domination as \cite{Caprini:2001nb}
\be
\label{eq_scrd}
a(\tau) = H_0\,\sqrt{\Omega_{\rm rad}}\,\tau\,,
\ee
and multiplying the final result by $\Omega_{\rm rad}$ to account for the redshifting of quantities evaluated during RD, we obtain
\bea
\label{eq_OGWB}
 \Omega_{\rm GW}^{(b)}\,=\,\frac{3}{128\,\pi^2} \,
\frac{\Omega_B^2 }{\Omega_{\rm rad}} 
 \left[\ln|k  {\tau_R}|  \right]^2\,F(k)\,.
\eea
The amplitude is consistent with~\cite{Caprini:2001nb}, up to numerical coefficients reflecting our focus here on contributions from scalar polarizations only.  Hence
in a setup based on scalar tensor theory we can expect two contributions
to the GW energy density  $\Omega_{\rm GW}$ -- one contribution $\Omega_{\rm GW}^{(b)}$ due to scalar modes, see Eq.~\eqref{eq_OGWB} -- the 
other controlled by tensors, with the same overall coefficient
${\Omega_B^2 }/{\Omega_{\rm rad}} $
  but different
numerical  coefficients and distinct dependence on momentum $k$. 

\smallskip
The methods developed in this preparatory section --  which computed
for the first time the energy density in GW scalar polarization 
as induced by magnetic field source -- 
set the stage for the  computation of the three-point function, which we now turn to.

 \subsection{The gravitational wave three-point function}
\label{sec_tpfgw}

We now compute   the  three-point function of the Fourier
modes  of the scalar polarization of gravitational waves, Eq.~\eqref{eq_frh}, sourced at second order
by the primordial magnetic field.  As we learned, such three point function
can constitute a particular transparent quantity to measure, being not 
contaminated
by spin-2 and spin-1 polarizations. 
The computation is conceptually  similar to the two-point function studied in the previous section. 
However, the new and interesting result we aim to highlight is that 
\emph{only folded configurations in momentum space}---corresponding to stationary signals in real space---lead to observable effects. See Fig \ref{fig_fold} for a representation
of folded triangles. All other
shapes of non-Gaussianity are washed out by time 
integrations involving highly oscillatory functions \cite{Bartolo:2018rku}. 
This illustrates the central role played by the stationary condition on the SGWB statistics in determining the measurability of the signal. 

The computation of the tensor three-point function depends on the
kind of interactions we wish to consider in the third order
action for tensor modes. As representative of modified 
gravity setup, we consider cubic terms involving time-derivatives
of $h_{ij}$, proportional 
to $\dot{h}_{ij}\dot{h}_{jk}\dot{h}_{ki}$, which can arise 
in theories containing Horndeski  interactions and are absent in General Relativity
(see e.g. \cite{Ozsoy:2019slf} for a cosmological 
application of such terms). Passing to conformal time and to Fourier modes,
and normalizing to make the quantities dimensionless, we
consider the following expression
for the bispectrum~\footnote{We indicate
with a hat the bispectrum, to differentiate
this definition with Eq.~\eqref{def_bis} where the
three-point function for quantities
without time derivatives is considered.} of the scalar polarization as
\begin{align}
\hat{B}_{b}({\bf k_1}, {\bf k_2}, {\bf k_3})\,&\equiv\, \frac{k_1 k_2 k_3}{(a H)^3}\langle
 h_{\bf k_1}(\tau) 
 h_{\bf k_2}(\tau) 
 h_{\bf k_3}(\tau) 
\rangle'_{\bf k_1+\bf k_2 +\bf k_3=0}
\nonumber \\
&= \,\frac{k_1 k_2 k_3}{(a^2\,H)^3}\,{\cal I}_{(3)}(\tau)\;
\langle
{S}_{\bf k_1}
{S}_{\bf k_2}
{S}_{\bf k_3}
\rangle'_{\bf k_1+\bf k_2 +\bf k_3=0} \,,
\label{eq_ts3s}
\end{align}
where the prime indicates that the overall momentum-conserving delta function has been factored out, and
\be\label{def_I3}
{\cal I}_{(3)}(\tau)\,\equiv\,\int d \tau_1 \, d \tau_2 \,d \tau_3 \;
\frac{g_{ k_1}(\tau, \tau_1)}{a(\tau_1)}\,
\frac{g_{ k_2}(\tau, \tau_2)}{a(\tau_2)}\,
\frac{g_{ k_3}(\tau, \tau_3)}{a(\tau_3)} \,.
\ee

\bigskip
\noindent
{\bf Time integral.}
The first main ingredient of Eq.~\eqref{eq_ts3s} is the time integral of
Eq.~\eqref{def_I3}, 
subject to the momentum-conservation condition ${\bf k_1+\bf k_2 + \bf k_3} = 0$. 
As in the two-point function analysis of Section \ref{sec_GWtpf}, the integrand is highly oscillatory.  
However, in this case the oscillations always involve odd combinations of sine and cosine terms, such as 
$\cos\!\big[(k_1+k_2+k_3)\tau\big]$, $\cos\!\big[(k_1+k_2-k_3)\tau\big]$, and so on.  
These average to zero unless the argument of the cosine vanishes.  
This occurs precisely for folded triangles, e.g.\ when $k_3=k_1+k_2$.  See Fig~\ref{fig_fold} for a graphical
representation.

For such configurations, the integral develops a logarithmic enhancement at late times.  
In the case $k_1+k_2=k_3$, we obtain
\bea
{\cal I}_{(3)}(\tau) &=& \frac{\pi}{8\,k_1 k_2 k_3}\,\frac{1}{H^3 a^6}\,\left[
\ln{(k_1 \tau_R)} \ln{(k_3 \tau_R)} +\ln{(k_2 \tau_R)} \ln{(k_3 \tau_R)} -\ln{(k_1 \tau_R)} \ln{(k_2 \tau_R)} 
\right]\nonumber
\\
&\equiv&\frac{1}{k_1 k_2 k_3\,(a^2 H)^3}\,G_2(k_1,k_2)\,.
\label{res_ti3p}
\eea
where we introduced the function $G_2$ of $k_1$, $k_2$, $k_3=k_1+k_2$
to keep track of the log-enhanced contributions.
The other two cases, $k_1+k_3=k_2$ and $k_2+k_3=k_1$, can be treated analogously.

\bigskip
\noindent
{\bf Source contribution.}
The last coefficient in Eq.~\eqref{eq_ts3s} -- the source three-point function -- can be obtained 
by carefully performing the convolution integrals. The result is
\[
\begin{aligned}
&\langle
{S}_{\bf k_1}
{S}_{\bf k_2}
{S}_{\bf k_3}
\rangle'
=\frac{k_{1a}k_{1b}k_{2c}k_{2d}k_{3e}k_{3f}}{k_1^2 k_2^2 k_3^2}
\langle B_a(\mathbf{p}_1) B_b(\mathbf{k}_1 - \mathbf{p}_1) B_c(\mathbf{p}_2) B_d(\mathbf{k}_2 - \mathbf{p}_2) B_e(\mathbf{p}_3) B_f(\mathbf{k}_3 - \mathbf{p}_3) \rangle' \\[6pt]
&\quad =\;
2\,\int \frac{d^3 \mathbf{p}_1}{(2\pi)^3}   P_B({p}_1) P_B(|\mathbf{k}_1 - \mathbf{p}_1|)
\Bigg\{
P_B(|\mathbf{k}_2 + \mathbf{p}_1|) \, \mathcal{T}_2(\mathbf{k}_1, \mathbf{k}_2; \mathbf{p}_1)
\;+\;
P_B(|\mathbf{k}_3 + \mathbf{p}_1|) \, \mathcal{T}_3(\mathbf{k}_1, \mathbf{k}_2; \mathbf{p}_1)
\Bigg\},
\end{aligned}
\]
where
\[
\begin{aligned}
\mathcal{T}_2(\mathbf{k}_1, \mathbf{k}_2; \mathbf{p}_1)
&= 
A\big(\mathbf{k}_1, \mathbf{k}_2; \mathbf{p}_1 \big) \,
A\big(\mathbf{k}_1, \mathbf{k}_3; \mathbf{k}_1 - \mathbf{p}_1 \big) \,
A\big(\mathbf{k}_2, \mathbf{k}_3; \mathbf{k}_2 + \mathbf{p}_1 \big) , 
\\[6pt]
\mathcal{T}_3(\mathbf{k}_1, \mathbf{k}_2; \mathbf{p}_1) 
&= A\big(\mathbf{k}_1, \mathbf{k}_3; \mathbf{p}_1 \big) \,
A\big(\mathbf{k}_1, \mathbf{k}_2; \mathbf{k}_1 - \mathbf{p}_1 \big) \,
A\big(\mathbf{k}_2, \mathbf{k}_3; \mathbf{k}_3 + \mathbf{p}_1 \big).
\end{aligned}
\]
The function $A$ appearing above is defined in Eq.~\eqref{def_fA}.  

\smallskip
\noindent
{\bf Folded configurations.} 
As we learned above, the physically  interesting limit is that of \emph{folded triangles}, in which the three momenta are collinear. See Fig \ref{fig_fold}. One example is
\be
{\bf k_1} \,=\,k_1 \,\hat a \,,\qquad {\bf k_2} \,=\,k_2 \,\hat a \,,\qquad {\bf k_3} \,=\,-k_3
\hat a \,,
\ee
with the condition $k_1+k_2=k_3$, for a
certain side direction $\hat a$. 
In this case it is convenient to define
\be
\mathcal{T}(\mathbf{u}) \;\equiv\; 1 - \frac{(\hat a \!\cdot\!\mathbf{u})^2}{|\mathbf{u}|^2}\,,
\ee
so that the source correlator reduces to
\[
\begin{aligned}
&\langle
{S}_{\bf k_1}
{S}_{\bf k_2}
{S}_{\bf k_3}
\rangle' 
= 2\,P_0^3
\int\frac{d^3\mathbf{p}_1}{(2\pi)^3}\;f({p}_1)\,f(|\mathbf{k}_1-\mathbf{p}_1|)\\[6pt]
&\qquad\times\Big\{
3\,f(|\mathbf{k}_2+\mathbf{p}_1|)\;
\mathcal{T}(\mathbf{p}_1)\,\mathcal{T}(\mathbf{k}_1-\mathbf{p}_1)\,\mathcal{T}(\mathbf{k}_2+\mathbf{p}_1) \\[4pt]
&\qquad\quad
+\;f(|\mathbf{k}_3+\mathbf{p}_1|)\;
\mathcal{T}(\mathbf{p}_1)\,\mathcal{T}(\mathbf{k}_1-\mathbf{p}_1)\,\mathcal{T}(\mathbf{k}_3+\mathbf{p}_1)
\Big\},
\\[6pt]
&\qquad\quad \equiv\, P_0^3 \,G_1(k_1,k_2, \hat a)
\label{res_s3pt}
\end{aligned}
\]
where the function $G_1$ in the last line of the previous formula 
depends on $k_1$, $k_2$, $k_3=k_1+k_2$, as well as on the direction $\hat a$
of the folded 
triangle sides in momentum space. 

\medskip

Substituting the results of Eqs~\eqref{res_s3pt} and \eqref{res_ti3p} in the definition
of the bispectrum \eqref{eq_ts3s}, using the expressions for
$\Omega_B$ and $a(\tau)$
 introduced in Section \ref{sec_GWtpf}, and multiplying the result by $\Omega_{\rm rad}$ to take
into account of redshifting during RD, we obtain

\be
\hat B_{b}({\bf k_1}, {\bf k_2}, {\bf k_3})\,=\,\frac{27\,\Omega_B^3}{\Omega_{\rm rad}^2}\, \left[ G_1(k_1,k_2, \hat a)\,G_2(k_1,k_2)+{\rm perms} \right]
\ee
and, due to considerations above, the previous 
quantity has support only in the folded limit, where the momenta
${\bf k}_i$ are all aligned.

\subsection{Summary of this section}
Our analysis shows that cosmological magnetic fields can source a non-trivial 
three-point function of scalar polarizations, which in principle constitutes a measurable 
signal. Such a detection would provide a powerful probe of alternative theories of gravity 
predicting extra polarizations beyond the tensorial modes of General Relativity. 

\smallskip
We find that the amplitude of the GW three-point function scales as 
\(
\Omega_B^3 / \Omega_{\rm rad}^2
\),
in contrast to the two-point function, whose amplitude scales as 
\(
\Omega_B^2 / \Omega_{\rm rad}
\).
The bispectrum can be further enhanced by  specific forms of the magnetic field spectrum $P_B(k)$, which enters the convolution integrals and may preferentially amplify the three-point signal on certain scales relative to the power spectrum. 
In addition, the two- and three-point functions acquire distinct logarithmic contributions from time integrals, which in some regions of parameter space can  increase non-Gaussian effects. A particularly illustrative case is provided by folded configurations, with momenta satisfying $k_1 \ll k_2, k_3$ and $k_1 + k_2 = k_3$. In this limit, the logarithmic term $\ln (k_1 \tau_R)$ enhances the bispectrum amplitude relative to the logarithmic factors that govern the power spectrum.  
These considerations suggest that one may evade existing bounds on $\Omega_B$ derived from the GW energy density of two-point correlators discussed in Section~\ref{sec_GWtpf}, while still obtaining a sizeable three-point signal on selected scales. Exploring these possibilities in detail, both analytically and numerically, represents an interesting direction for future work, which we plan to pursue in forthcoming publications.

\smallskip

A key outcome of our computation is that, due to time-averaging over rapid oscillations, 
only bispectrum shapes corresponding to \emph{folded configurations}---which describe 
stationary processes in real space---yield non-vanishing contributions. 
All other configurations are suppressed: their contributions average to zero because of 
decorrelation effects, as first emphasized in \cite{Bartolo:2018rku}.  

The results of this section provide a concrete demonstration of the central role of the stationary condition 
for the feasibility of detecting scalar polarizations with our method.  If scalar polarizations
exist, they can be generated at second order in perturbations by sources 
as primordial magnetic fields, as examined here. 
They can contribute
to the total GW two-point function -- as studied in Section \ref{sec_GWtpf} -- hence 
contaminating measurements of the GW energy density. More interestingly
to us, they can source  three-point functions with support in folded limits, which
if measured provide a unique footprint of modified gravity.  More in general,
we expect similar effects to occur in systems where amplified fluctuations -- in
scalar or vector sectors -- source GW at second order in perturbations.

\section{Conclusions}
\label{sec_concl}

In this work we have proposed a novel method to detect scalar polarizations in the stochastic gravitational-wave (GW) background, as predicted in theories beyond General Relativity. Our approach is based on measuring three-point correlation functions of the GW signal. After averaging over the polarization angle -- which accounts for the ambiguity in the definition of polarization tensors -- we have shown that the detector response to three-point correlations vanishes for spin-2 and spin-1 modes, but remains non-vanishing for scalar modes. Therefore, the detection of a non-zero GW three-point correlator would provide direct evidence for the existence of scalar polarizations. 

We derived analytical expressions for the response functions to GW three-point correlators for coincident ground-based detectors, for pulsar timing arrays (PTAs), and for astrometric measurements. Our analysis included the possibility of cross-correlating different observables, and we characterized the geometrical properties of the corresponding response functions. Furthermore, we constructed an optimal estimator for the GW three-point function and developed simple Fisher forecasts to assess the detectability of its amplitude with PTA experiments. As an illustration, we presented a concrete cosmological scenario in which GWs are induced at second order in a primordial magnetogenesis setup, showing that such a mechanism can in principle generate a sizeable three-point signal. 

Future work will involve refining and extending the early-universe model we introduced, as well as applying our formalism to current and forthcoming GW datasets. This will allow us to place explicit bounds on the amplitude of GW three-point correlations and, crucially, to constrain or discover the presence of scalar polarizations in the GW background.

\subsection*{Acknowledgements}
It is a pleasure to thank
Ameek Malhotra for discussions.
We are partially funded by the STFC grants ST/T000813/1 and ST/X000648/1.  For the purpose of open access, the authors have applied a Creative Commons Attribution licence to any Author Accepted Manuscript version arising.

{\small
\providecommand{\href}[2]{#2}\begingroup\raggedright\endgroup

}

\end{document}